\documentclass{ws-mpla}

\begin{document}

\markboth{A.~Meyer}
{First CDF Run~II Results}

%
\catchline{}{}{}{}{}
%

\title{THE CDF EXPERIMENT AT THE TEVATRON --- \\
THE FIRST TWO YEARS OF RUN~II\footnote{Talk given at the DESY ``Tuesday seminar'' in
November 2002}}

\author{\footnotesize ARND MEYER
}

\address{III. Physikalisches Institut A, RWTH Aachen, Physikzentrum \\
52056 Aachen, Germany \\
meyera@fnal.gov}

\author{For the CDF Collaboration}

\maketitle


\begin{abstract}
The 1992-1995 running of the Fermilab Tevatron (the so-called Run~I) ended with
many important physics goals accomplished, including the discovery of the top
quark, and the anticipation of many further questions to be answered in the
future. After many upgrades to the detector and to the accelerator complex, Run~II
began in April 2001. First results obtained by the Collider Detector at
Fermilab (CDF) collaboration from the analysis of early Tevatron Run~II data are
reported here. They fall in two categories: a number of measurements have been
performed with the primary goal of establishing detector performance and physics
potential. Another set of measurements make use of completely new capabilities
of the upgraded detector, thus allowing for competitive results with a modest
amount of integrated luminosity.

\end{abstract}

\ccode{PACS Nos.: 13.25.Hw, 13.85.-t, 13.85.Ni, 13.85.Qk, 13.85.Rm, 13.87.-a,
                  14.20.Mr, 14.40.Lb, 14.40.Nd, 14.65.Ha, 14.70.Fm, 14.70.Hp}

\section{Introduction}

  There have been numerous tantalizing suggestions, both experimental and
  theoretical, that physics beyond the Standard Model might be within our
  reach. After the very successful Tevatron Run~I (1992-1995), colliding
  protons and antiprotons with a center of mass energy of $\sqrt s =
  1.8\:\mbox{TeV}$, the accelerator and the colliding beam experiments CDF and
  D\O\ were upgraded for Run~II, in order to expand the reach toward whatever new
  physics might exist.

  On the accelerator side, the beam energy has been increased from
  $900\:\mbox{GeV}$ to $980\:\mbox{GeV}$. For some rare processes like
  $t\bar{t}$ production this modest increase translates into a cross section
  higher by 30\% or more. In Run~I the luminosity reached $1.5\cdot
  10^{31}\:\mbox{cm}^{-2}\mbox{s}^{-1}$, while the ultimate goal for Run~II is
  $2-4\cdot 10^{32}\:\mbox{cm}^{-2}\mbox{s}^{-1}$. A new proton storage ring,
  the ``Main Injector'', has been built, and the same tunnel houses the
  ``Recycler'' storage ring, which will improve the rate at which antiprotons
  can be accumulated, and may also be utilized to reuse antiprotons after
  recovering them from the Tevatron. The bunch spacing in the Tevatron has been
  reduced from $3.5\:\mu\mbox{s}$ to $396\:\mbox{ns}$, corresponding to 36
  colliding proton and antiproton bunches, mostly to keep the number of
  overlapping events in the experiments small.

\begin{figure}[tb]
   \setlength{\unitlength}{1cm}
   \begin{picture}(15.0,3.6)(0.0,0.0)
   \put(-0.1,-0.8){\psfig{file=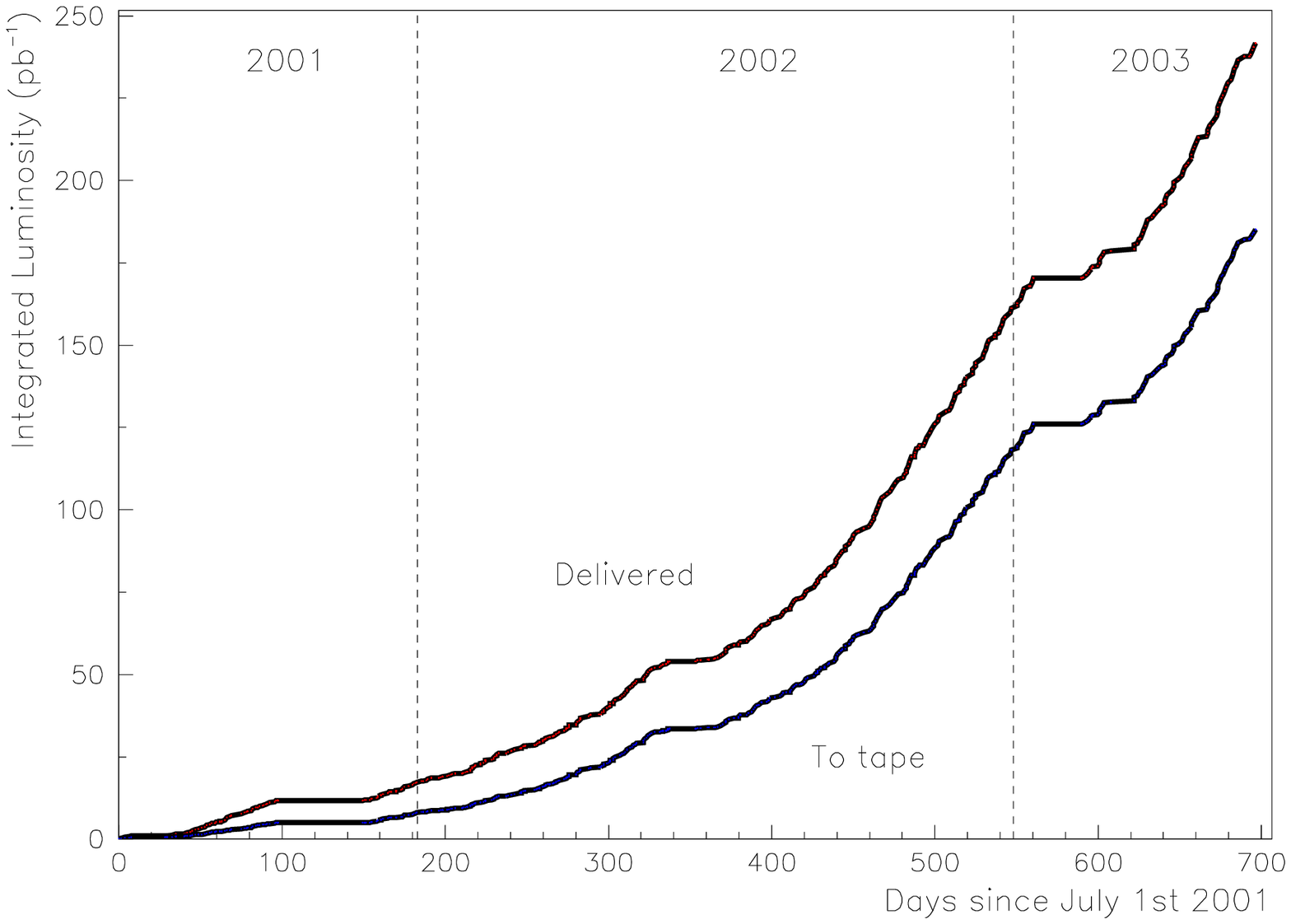,width=6.5cm}}
   \put(6.4,-0.6){\psfig{file=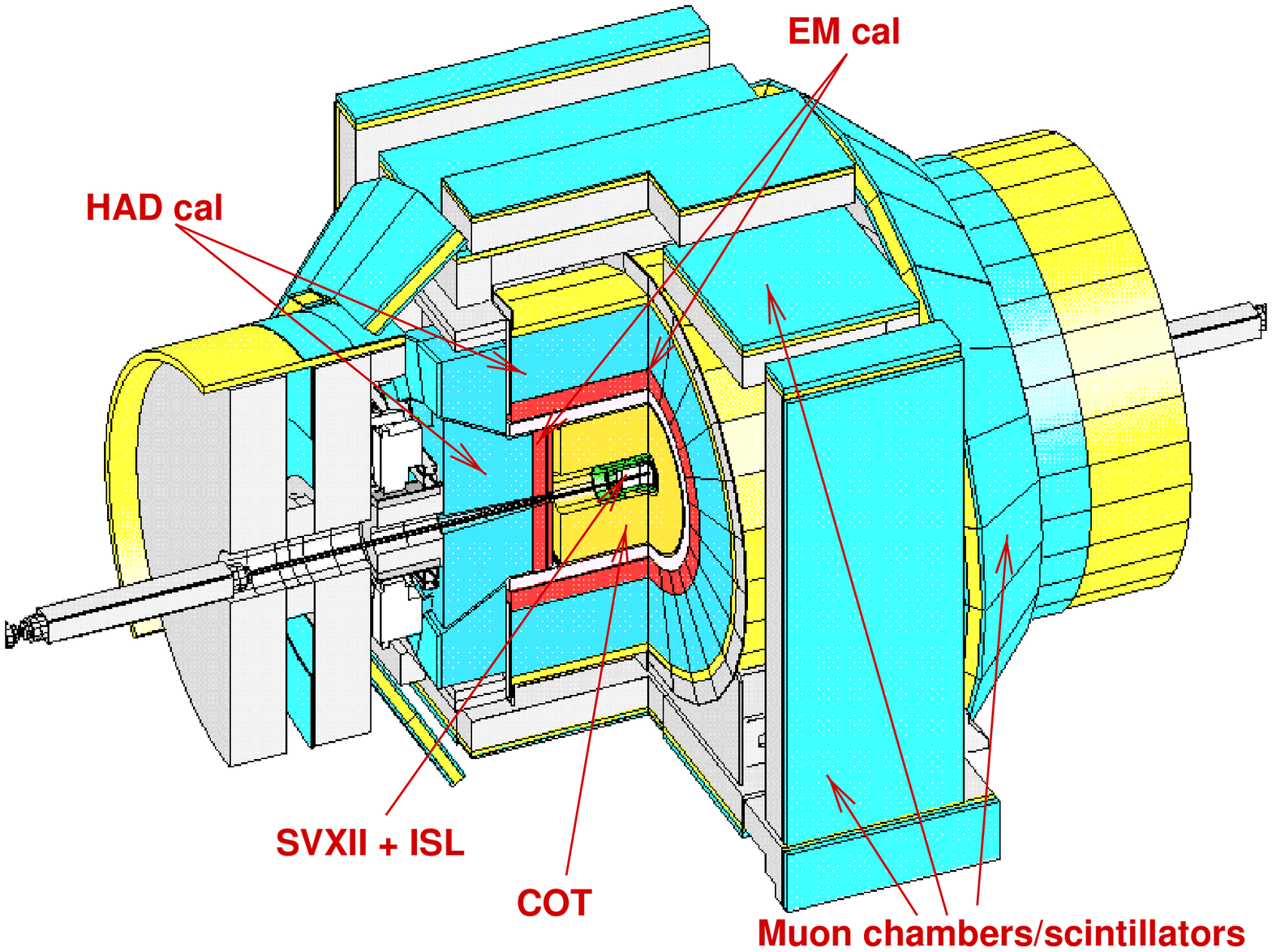,width=6.5cm}}
   \end{picture}
   \vspace*{8pt}
   \caption{{\em Left:} Delivered and recorded luminosity for CDF during early Run~II.
             {\em Right:} Schematic of the CDF Run~II detector.}
   \label{cdf}
  \end{figure}

  By June 2003, the Tevatron routinely achieved peak luminosities of $4-5\cdot
  10^{31}\:\mbox{cm}^{-2}\mbox{s}^{-1}$. The integrated luminosity delivered to
  CDF between July 2001 and June 2003 amounts to $240\:\mbox{pb}^{-1}$, and CDF
  recorded about $185\:\mbox{pb}^{-1}$ (Fig.~\ref{cdf}). The detector has been
  fully functional since early 2002, and the results shown below are based on up
  to $85\:\mbox{pb}^{-1}$ collected mostly in 2002, a data sample comparable to
  the Run~I data set ($\int {\cal L}\, dt \simeq 110\:\mbox{pb}^{-1}$).

  Run~II will be split into two major periods of data taking: Run~IIa, which is
  scheduled to last until the end of 2005, should provide at least
  $2\:\mbox{fb}^{-1}$ of integrated luminosity. Run~IIb will last until the
  beginning of the LHC physics program, bringing the total data sample to about
  $10\:\mbox{fb}^{-1}$.

\section{Detector and Trigger}

\begin{figure}[tb]
   \setlength{\unitlength}{1cm}
   \begin{picture}(15.0,4.9)(0.0,0.0)
   \put(0.0,-0.6){\psfig{file=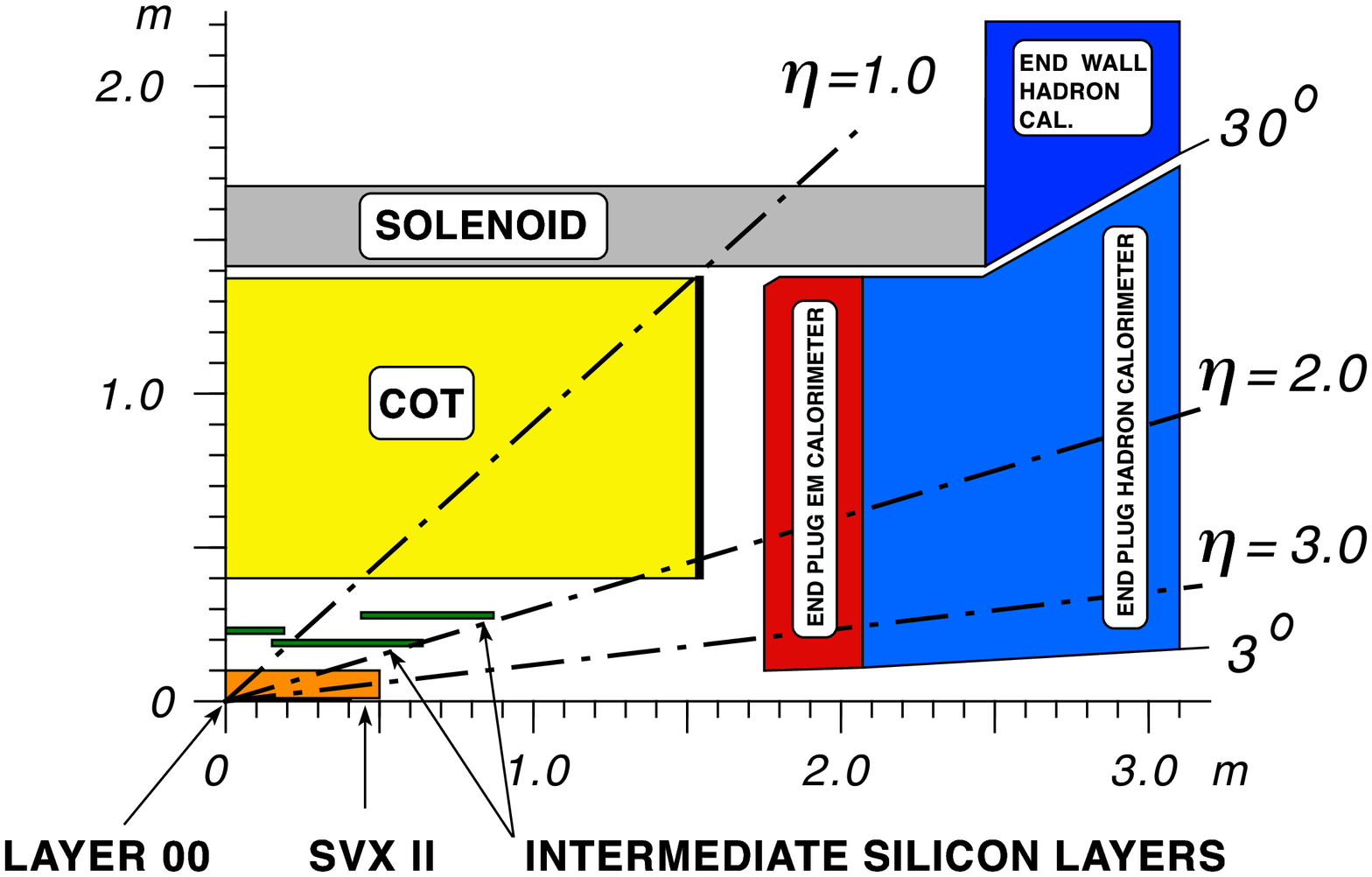,width=6.9cm}}
   \put(7.1,-0.8){\psfig{file=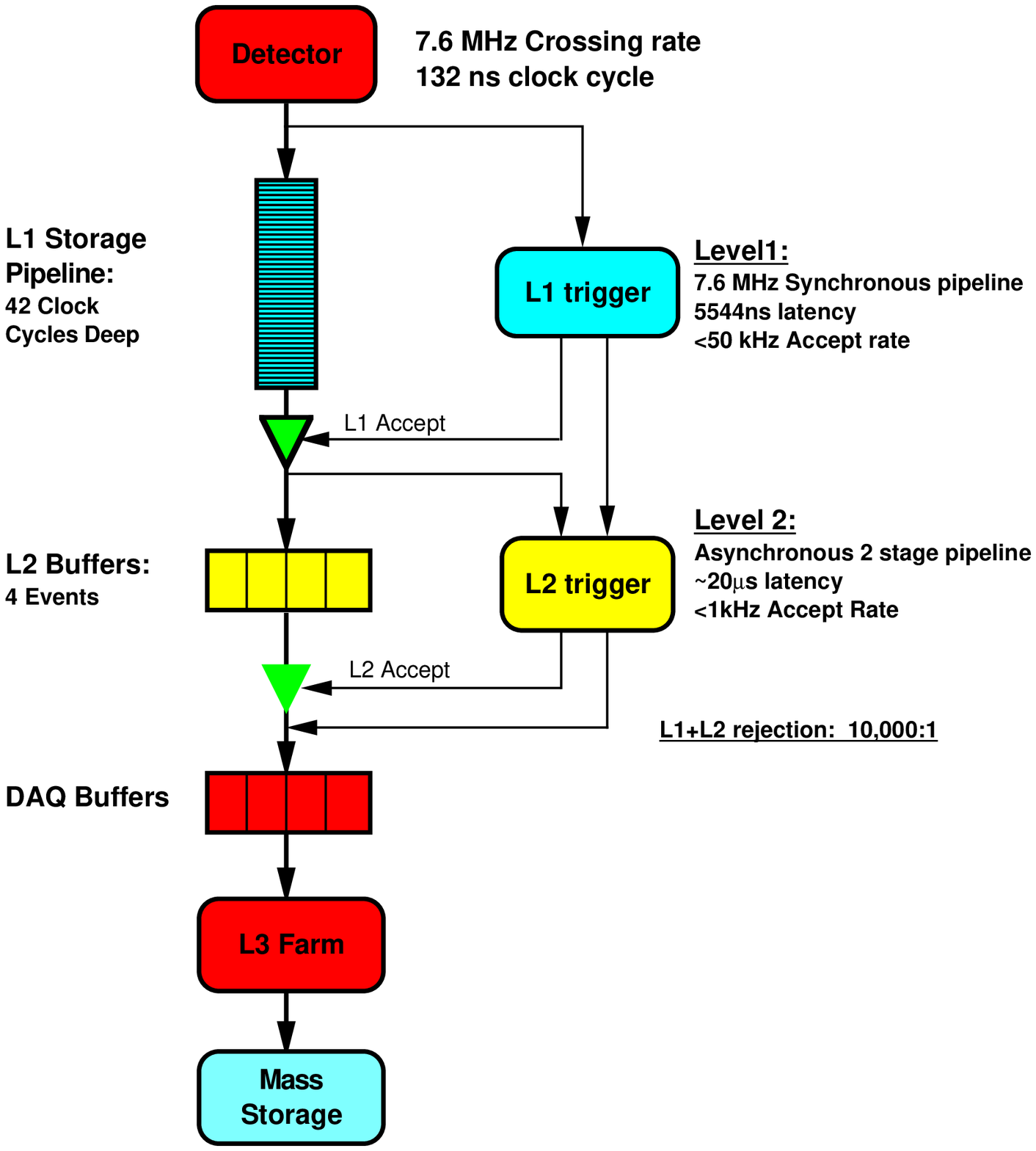,width=5.5cm}}
   \end{picture}
   \vspace*{8pt}
   \caption{{\em Left:} Side view of the tracking system.
            {\em Right:} The CDF trigger system for Run~II.}
   \label{cdftrack}
  \end{figure}

  The CDF detector (Fig.~\ref{cdf}) is designed for general purpose use, with a large
  tracking system inside a uniform $1.4\:\mbox{T}$ solenoidal magnetic field, $4\pi$
  calorimetry, and a muon detection system. A detailed description of the CDF upgrade
  project for Run~II can be found in Ref.~\refcite{TDR}. The entire tracking system
  (Fig.~\ref{cdftrack}) has been replaced for Run~II. The Silicon tracking system consists
  of three main components, such that a typical particle traverses seven layers of Silicon
  at radii of $1.5\:\mbox{cm}$ to $28\:\mbox{cm}$. Its main purpose is the precise
  measurement of secondary vertices from $b$ decays. Next, the Central Outer Tracker (COT)
  is an open cell drift chamber for the precise momentum measurement of charged tracks,
  using up to 96 space points, and with an efficiency of close to 100\% for high $p_T$
  isolated tracks. Between the COT and the solenoid is the new Time-of-Flight detector
  (TOF), with a time resolution of $100\:\mbox{ps}$ and consisting of 216 about
  $3\:\mbox{m}$ long scintillator bars, which together with the momentum measurement
  provides particle identification by determining a particle's mass. Electromagnetic and
  hadronic calorimeters are situated outside the solenoid. Electrons, photons, and jets
  deposit almost all their energy in the calorimeters. Muons travel through the
  calorimeters depositing only a small fraction of their energy, and are detected by the
  muon chambers which surround the calorimeters and steel absorbers.  The luminosity is
  measured with two modules of 48 low-mass Cherenkov counters, covering the rapidity
  region from $3.75$ to $4.75$, by counting the rate of inelastic $p\bar{p}$ interactions,
  and verified with measurements of well known physics processes like $W$ production.

  The three-level trigger system and the data acquisition system\cite{daq} have been
  significantly enhanced for Run~II (Fig.~\ref{cdftrack}). The Level 1 trigger is
  fully pipelined with a depth of $42\times 132\:\mbox{ns}$, mostly based on
  custom-built hardware, and delivers a decision every $132\:\mbox{ns}$. The
  following trigger levels are highly buffered, and the event digitization and
  readout does not interrupt triggering, so that the entire system is to first order
  deadtime-free. The first trigger level provides fast drift chamber tracks with high
  precision and efficiency, muon and electron triggers based on signatures in the
  muon chambers and calorimeter in combination with drift chamber tracks, and
  calorimeter based triggers looking for missing transverse energy, photons, and
  jets. The Level 1 accept rate achieved to date is about $20\:\mbox{kHz}$. At the
  second trigger level the rate is reduced with the help of refined information from
  all detector components. The major part of the rate reduction at Level 2 to about
  $300\:\mbox{Hz}$ is provided by the secondary vertex trigger, using information
  from the Silicon system. The third trigger level consists of a conventional farm of
  Linux PC's. A filter program based on an optimized version of the regular
  reconstruction program reduces the rate to typically $50\:\mbox{Hz}$ which are
  written to permanent storage (disk and tape).

\section{Charm and Beauty}

  Traditionally heavy flavor physics at hadron colliders relies on a lepton
  signature; examples are the decay of the $J/\psi$ into $\mu^+\mu^-$ or
  semileptonic $b$ decays. With the Silicon Vertex Tracker (SVT)\cite{SVT} CDF has
  introduced a novel method to obtain heavy flavor decays. The SVT uses COT tracks
  as seeds to a parallelized pattern recognition in the Silicon vertex detector. The
  following linearized track fit returns track parameters with nearly offline
  resolution on a time scale of $15\:\mu\mbox{s}$. The precise measurement of the
  track impact parameter ($\simeq 50\:\mu\mbox{m}$ including a $\simeq 35\:\mu\mbox{m}$
  contribution from the transverse size of the beams) makes it possible to trigger on
  displaced tracks from long-lived hadrons containing heavy flavor. Originally designed
  to select hadronic $B$ decays the SVT also collects a large sample of charm hadrons.

\subsection{Charm}

  All of the charm measurements shown here, with the exception of the $J/\psi$
  cross section, are based on data samples collected with the SVT.

\subsubsection{Prompt $D$ Meson Cross Sections}

\begin{figure}[tb]
   \setlength{\unitlength}{1cm}
   \begin{picture}(15.0,6.0)(0.0,0.0)
   \put(0.8,3.0){\psfig{file=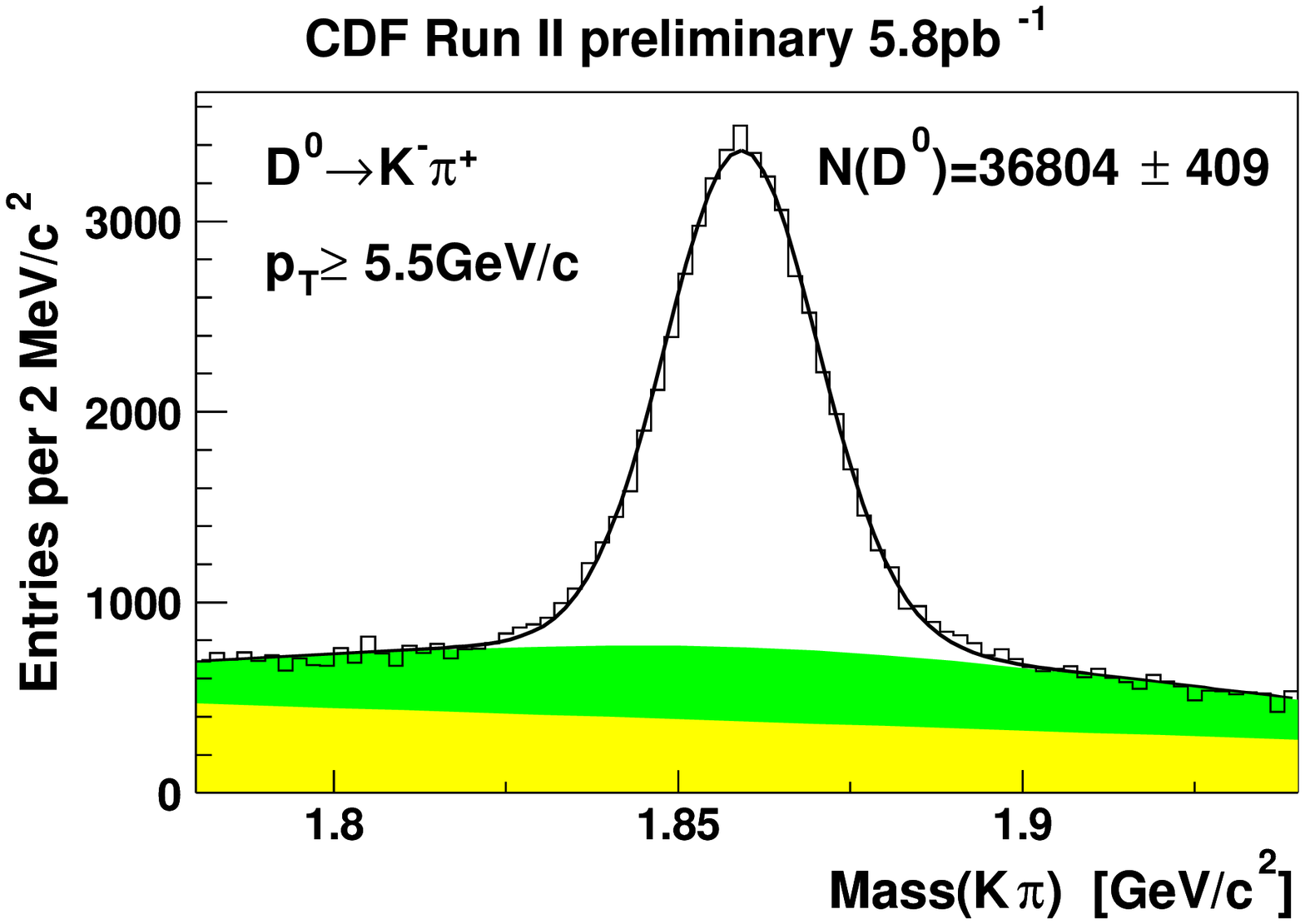,width=5.0cm}}
   \put(6.5,3.0){\psfig{file=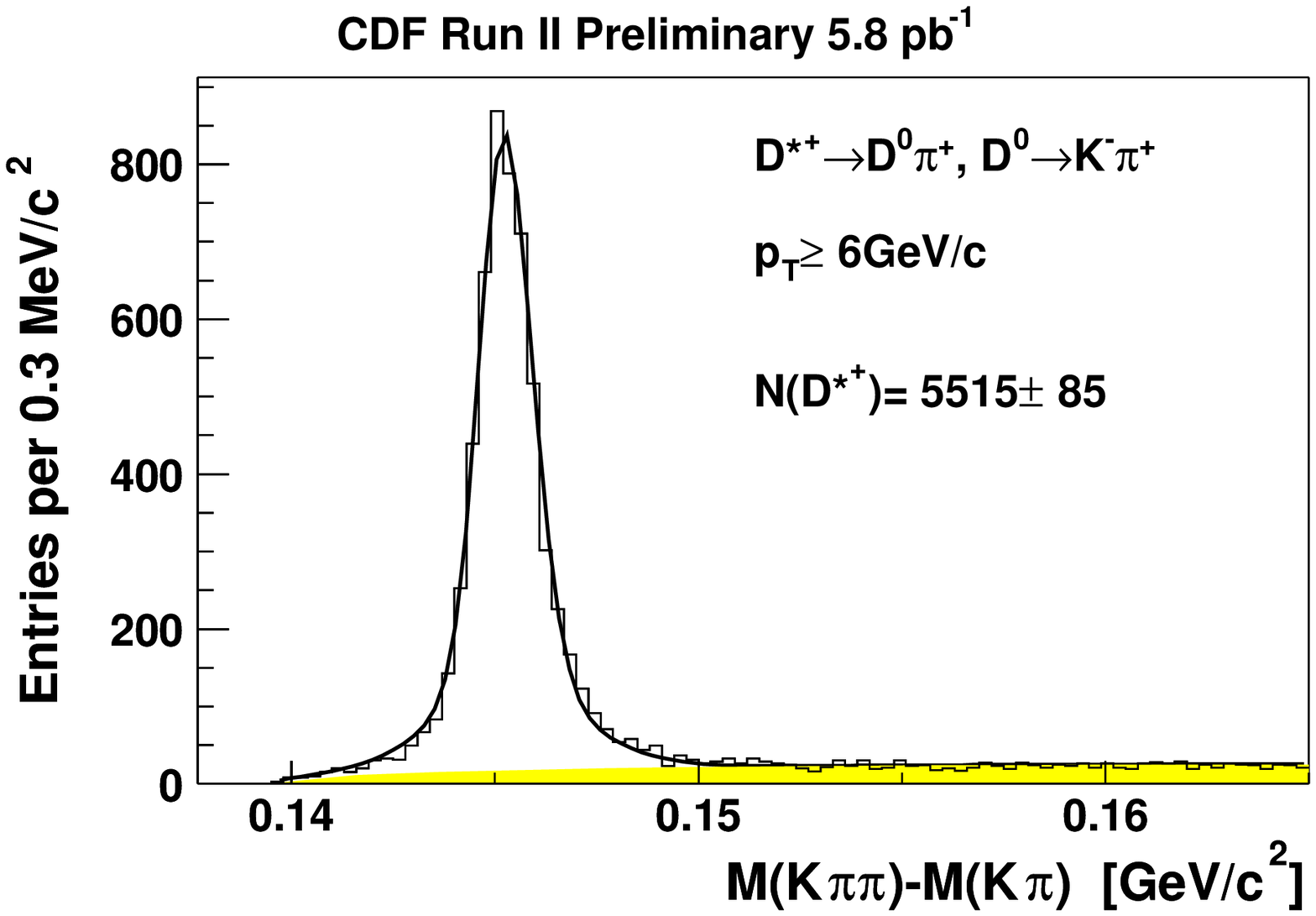,width=5.0cm}}
   \put(0.8,-0.8){\psfig{file=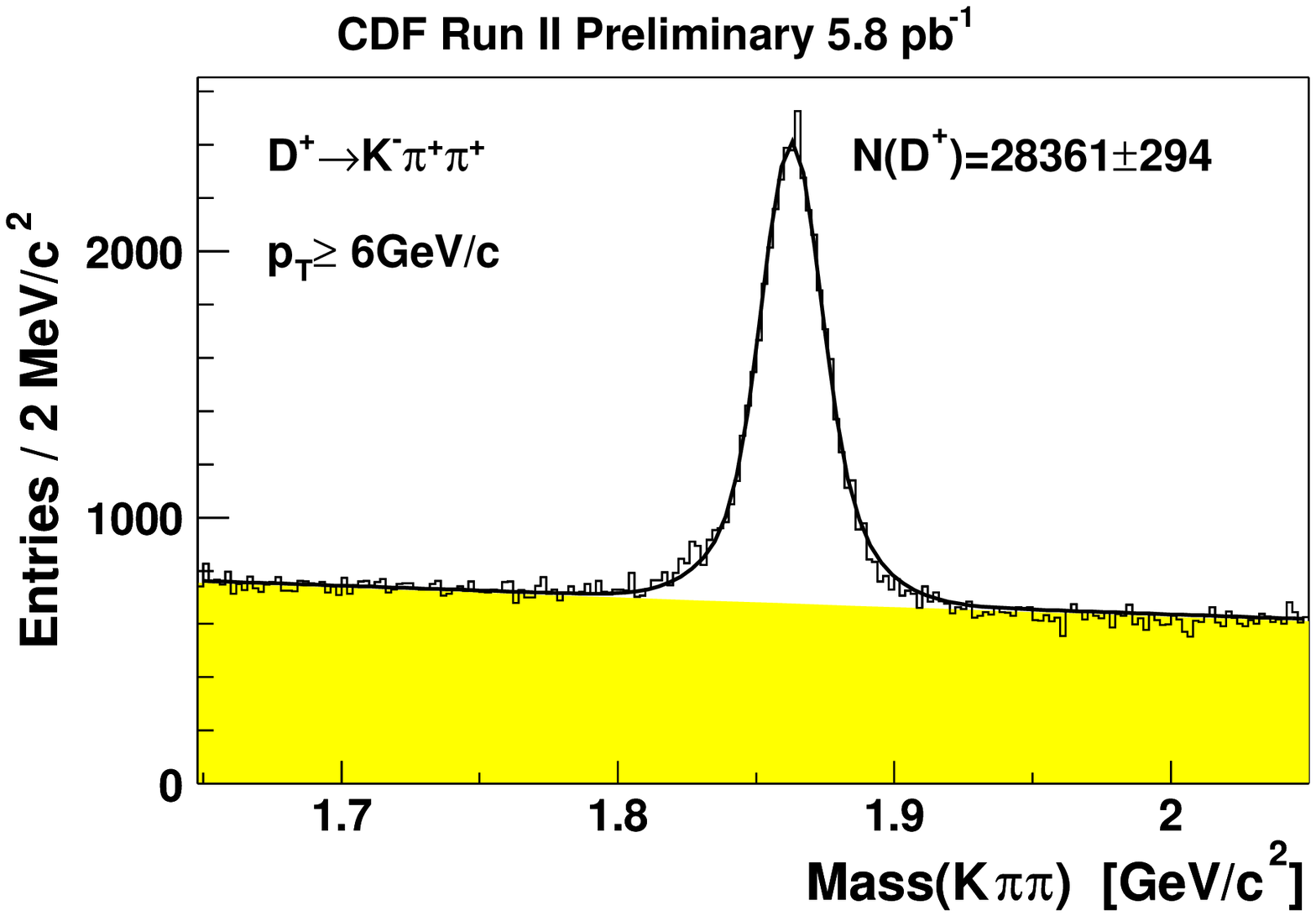,width=5.0cm}}
   \put(6.5,-0.8){\psfig{file=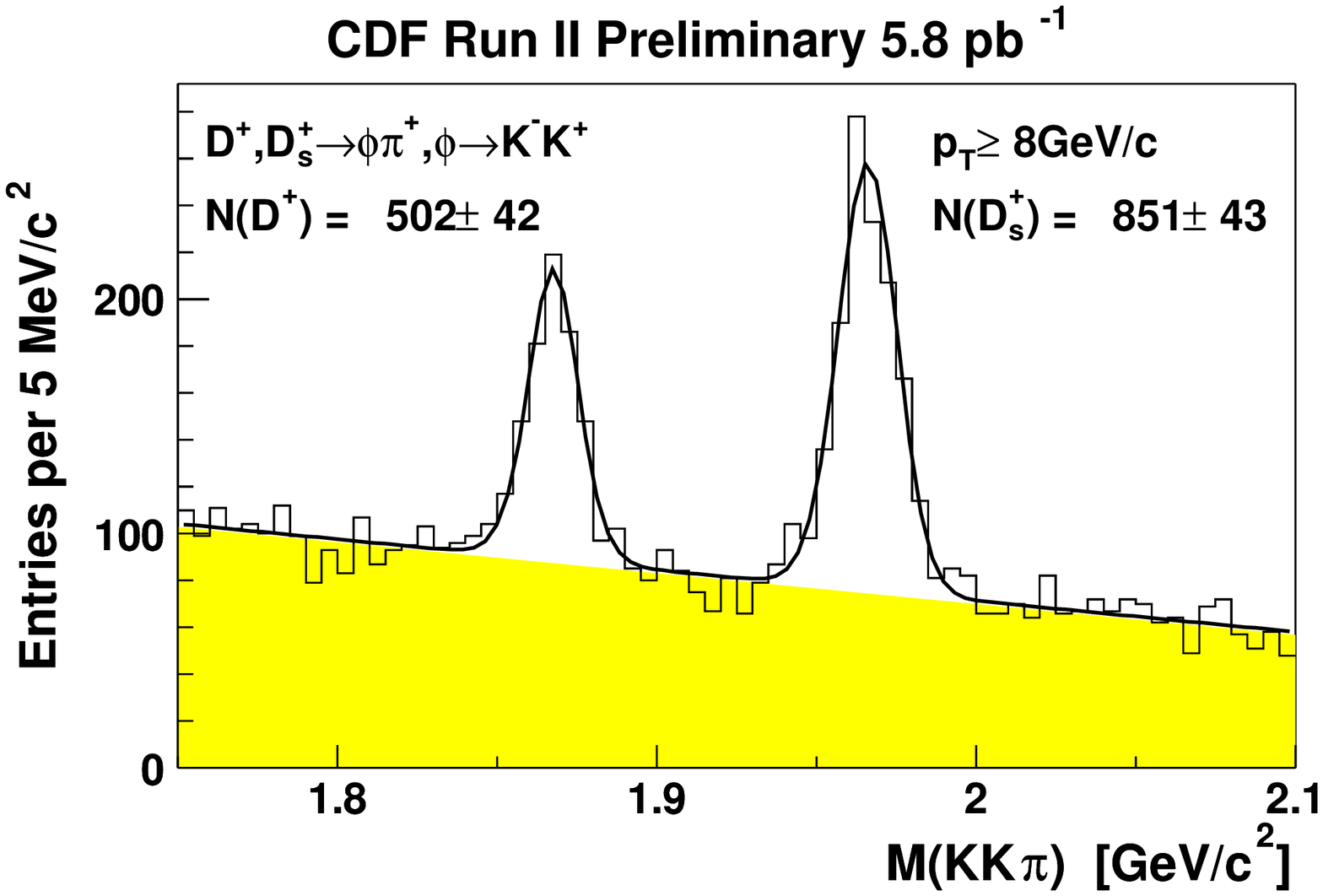,width=5.0cm}}
   \end{picture}
   \vspace*{8pt}
   \caption{Reconstructed mass spectra for the four reconstructed $D$ meson decays. The
            shaded area indicates the fitted combinatorial background. For the $D^0$
            the second shaded area shows the autoreflection contribution (exchange of
            $K$ and $\pi$). The signals are
            parameterized by single ($D^0$, $D_s^+$, $D^+ \rightarrow K^-\pi^+\pi^+$)
            or double ($D^{\ast +}$, $D^+\rightarrow \phi\pi^+$) Gaussians, respectively.}
   \label{csignals}
  \end{figure}

  There is no published measurement of the charm cross section from the Tevatron.
  With the advent of the SVT this measurement is possible in Run~II, and it is of
  theoretical interest due to the larger than expected beauty cross sections
  compared to next-to-leading order QCD calculations. The measurement shown here,
  based on the analysis of $5.8\:\mbox{pb}^{-1}$ of data, makes use of four fully
  reconstructed decay modes: $D^0 \rightarrow K^-\pi^+$, $D^{\ast +} \rightarrow
  D^0\pi^+$ with $D^0 \rightarrow K^-\pi^+$, $D^+ \rightarrow K^-\pi^+\pi^+$, and
  $D_s^+ \rightarrow \phi\pi^+$ with $\phi \rightarrow K^+K^-$. The mass spectra
  are shown in Fig.~\ref{csignals}.

  The contributions from prompt and secondary charm are separated by utilizing
  the impact parameter distribution of the reconstructed $D$ meson samples.
  Mesons originating from $B$ decays exhibit a large impact parameter. A fit to
  the impact parameter distribution yields prompt production fractions of
  $88.6\pm 0.4 \mbox{(stat)} \pm 3.5 \mbox{(sys)}$\%, $88.1\pm 1.1\pm 3.9$\%,
  $89.1\pm 0.4\pm 2.8$\%, and $77.3\pm 3.8\pm 2.1$\% for $D^0$, $D^+$, $D^+$, and
  $D_s^+$, respectively, averaged over the full analyzed $p_T$ range.

\begin{figure}[tb]
   \setlength{\unitlength}{1cm}
   \begin{picture}(15.0,7.0)(0.0,0.0)
   \put(0.2,3.3){\psfig{file=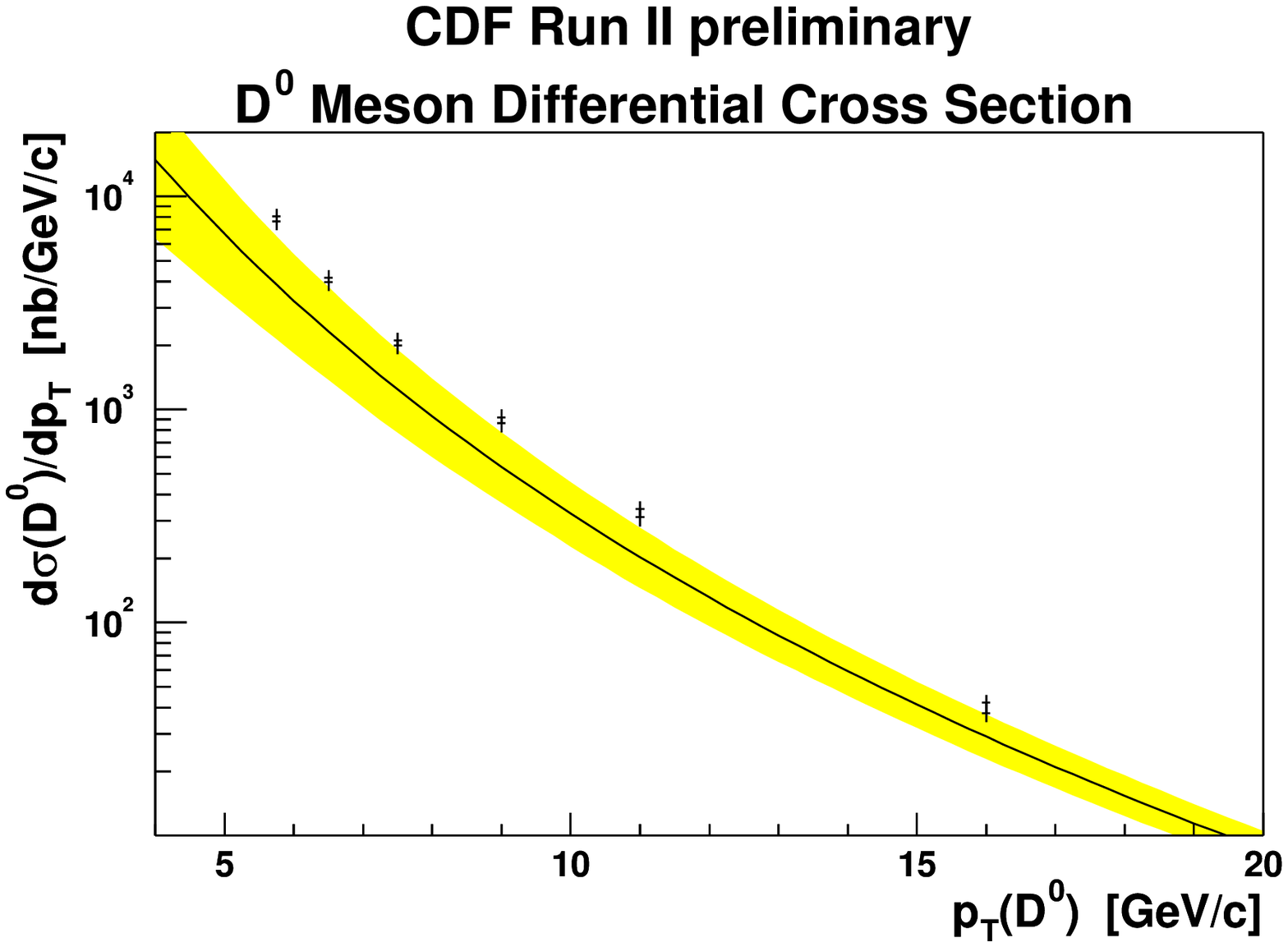,width=6.0cm}}
   \put(6.6,3.3){\psfig{file=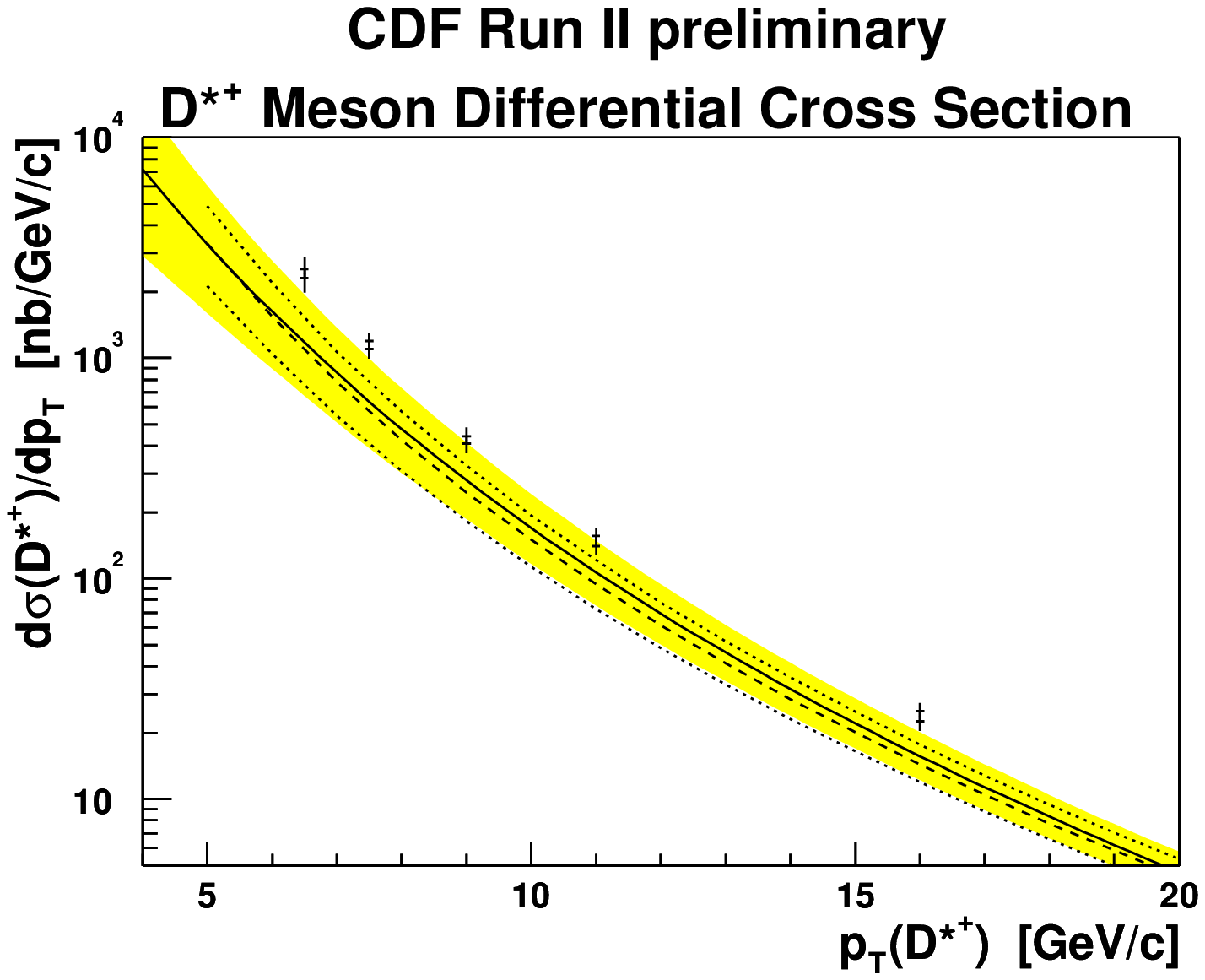,width=6.0cm,height=4.1cm}}
   \put(0.2,-0.8){\psfig{file=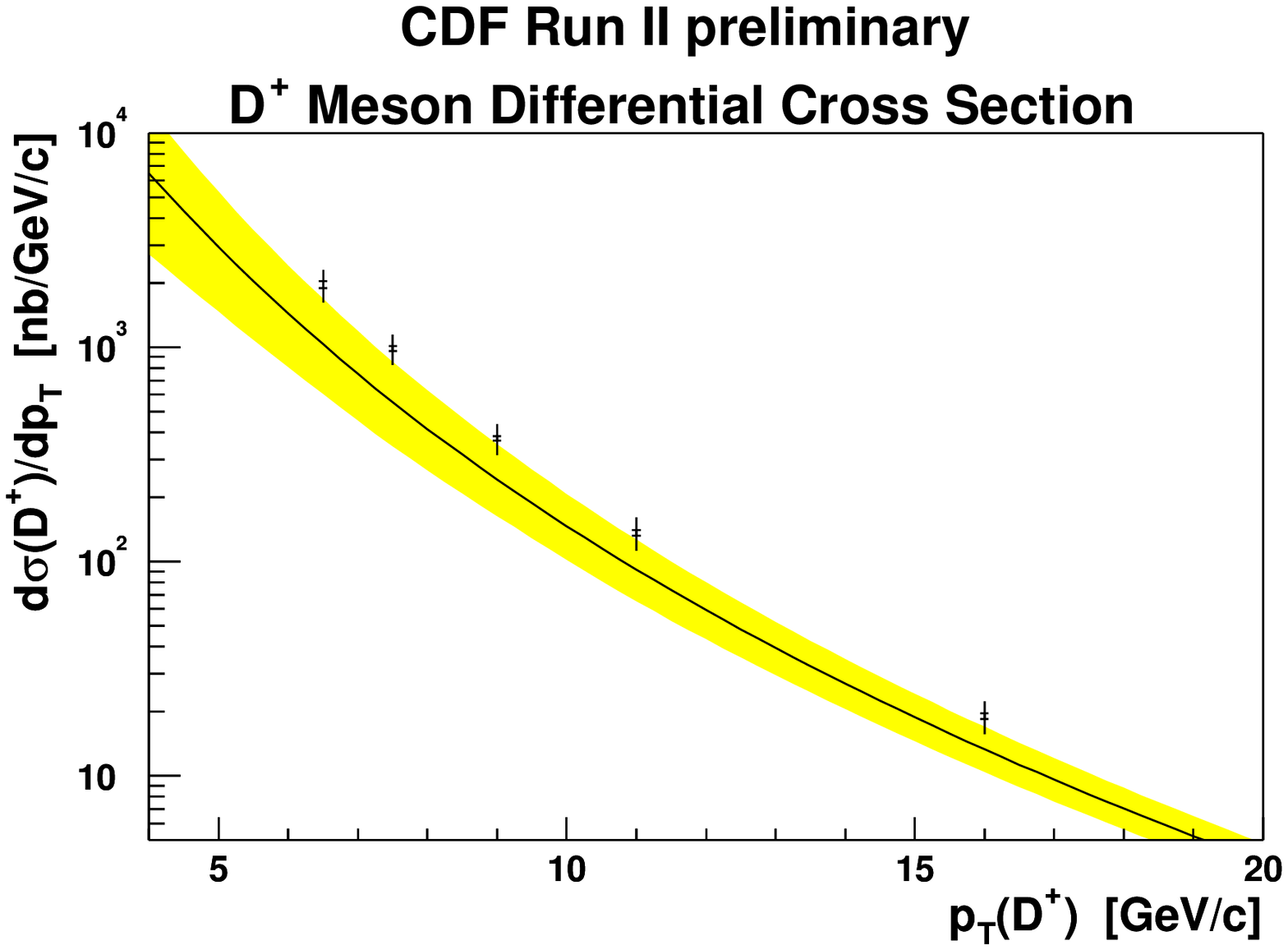,width=6.0cm}}
   \put(6.6,-0.8){\psfig{file=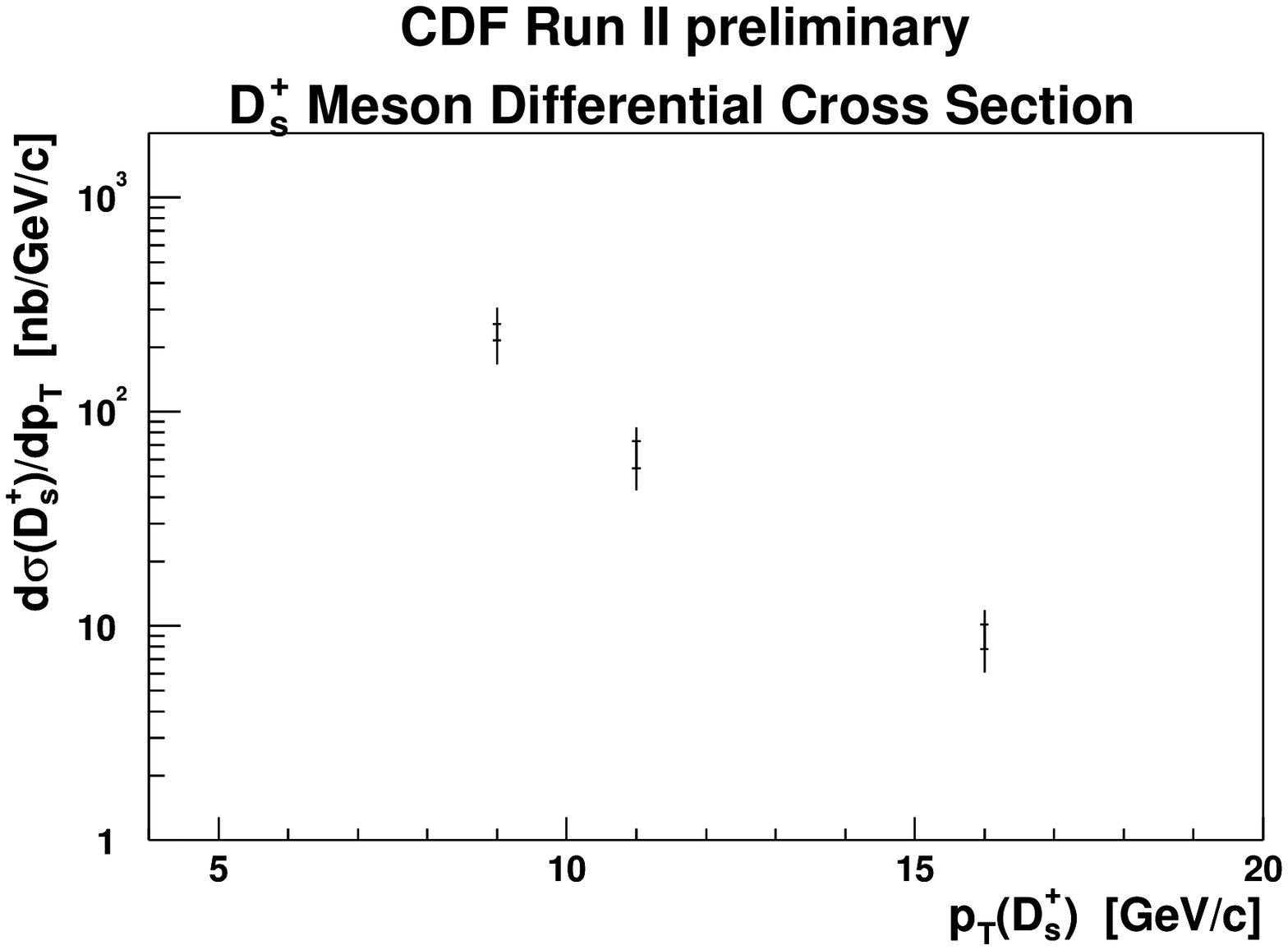,width=6.0cm}}
   \end{picture}
   \vspace*{8pt}
   \caption{Differential cross section for the four reconstructed $D$ mesons. Theoretical
            predictions from Kniehl {\em et al.} (dashed line with dotted line indicating
            uncertainty)
            and Cacciari {\em et al.} (full line with shaded band as uncertainty) are
            compared to the data. The theoretical uncertainties are based on varying the 
            renormalization and factorization scales independently by factors of 0.5 to 2.}
   \label{cxsec}
  \end{figure}

  The measured prompt differential cross sections are shown in Fig.~\ref{cxsec}.
  They are compared to a next-to-leading order QCD calculation\cite{kniehl}, and a
  fixed order next-to-leading log calculation\cite{fonll}. The calculations are
  lower than, but compatible with the data. The total cross sections are found to
  be:
  $\sigma (D^0, p_T>5.5\:\mbox{GeV})        = 13.3 \pm 0.2 \pm 1.5\:\mu\mbox{b}$,
  $\sigma (D^{\ast +}, p_T>6.0\:\mbox{GeV}) = 5.4 \pm 0.1 \pm 0.8\:\mu\mbox{b}$,
  $\sigma (D^+, p_T>6.0\:\mbox{GeV})        = 4.3 \pm 0.1 \pm 0.7\:\mu\mbox{b}$, and
  $\sigma (D_s^+, p_T>8.0\:\mbox{GeV})      = 0.75 \pm 0.05 \pm 0.22\:\mu\mbox{b}$, where
  the first error is statistical, and the second covers systematic uncertainties.

\subsubsection{$m_{D_s^+}-m_{D^+}$ Mass Difference}

  The measurement of the $m_{D_s^+}-m_{D^+}$ mass difference provides a test for
  the Heavy Quark Effective Theory and lattice QCD. While many precision measurements
  of meson masses can be expected from Run~II, the analysis shown here could
  already be carried out with a modest amount of luminosity.

\begin{figure}[tb]
   \setlength{\unitlength}{1cm}
   \begin{picture}(15.0,3.3)(0.0,0.0)
   \put(0.1,-0.9){\psfig{file=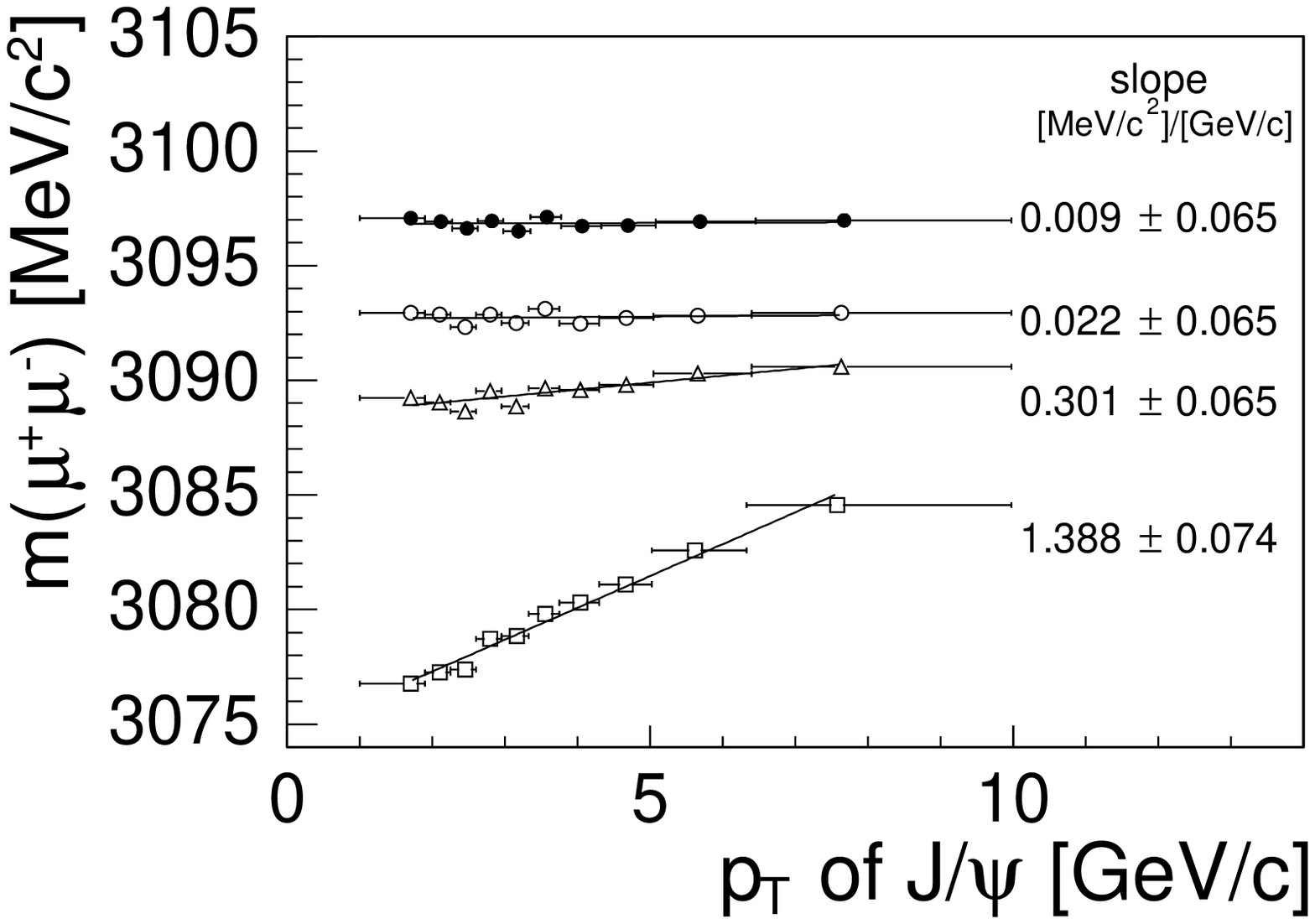,width=6.5cm}}
   \put(6.8,-0.9){\psfig{file=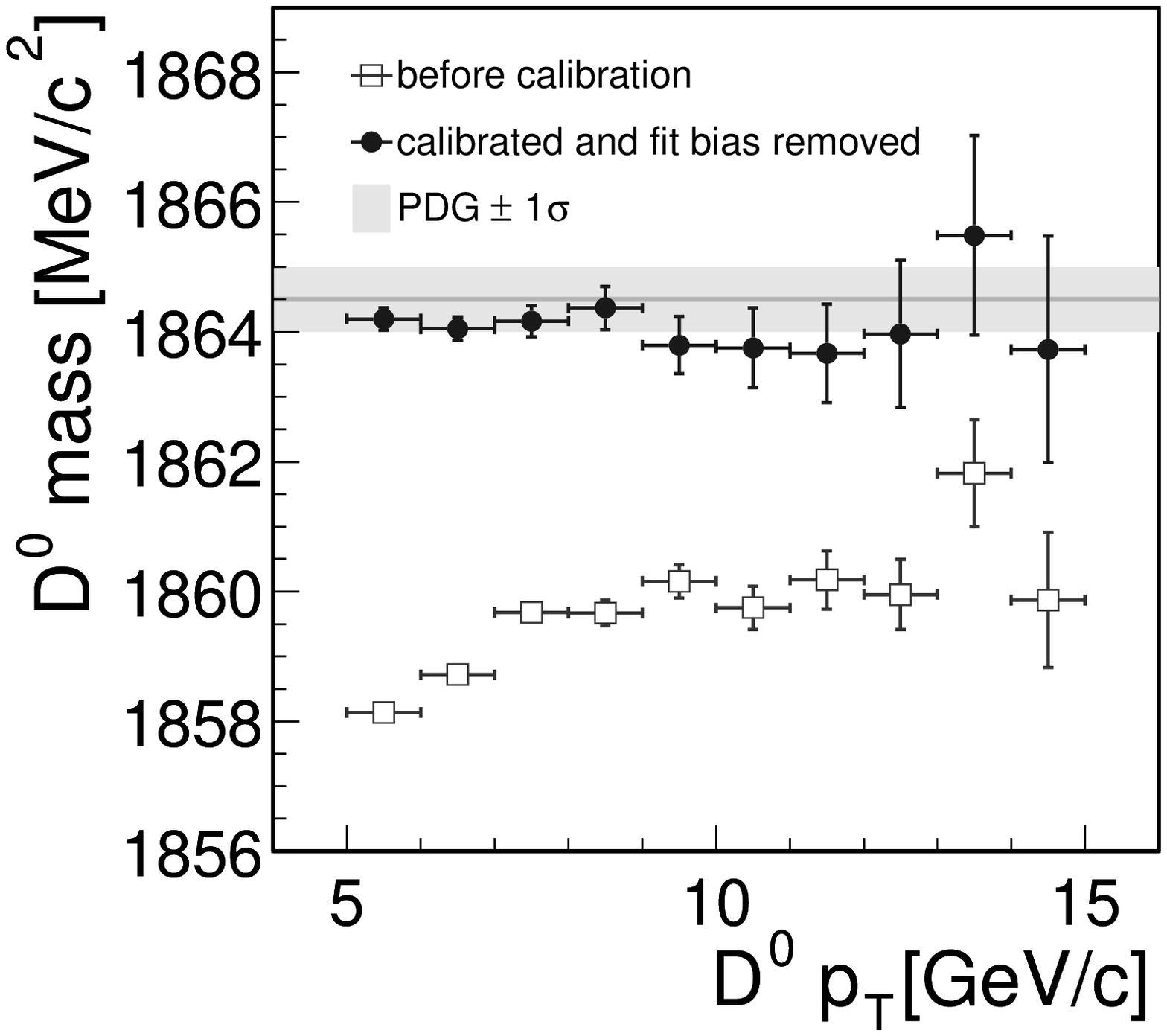,width=5.4cm}}
   \end{picture}
   \vspace*{8pt}
   \caption{Momentum scale calibration. {\em Left:} Illustration of the procedure using the
            momentum dependence of the reconstructed $J/\psi$ mass. The four sets
            of measurements are, from bottom to top: raw track reconstruction, inclusion
            of material effects, fine tuning of material, fine tuning of the magnetic
            field.
            {\em Right:} Validation with $D^0 \rightarrow K\pi$ decays.}
   \label{masscalib}
  \end{figure}

  For all mass measurements a calibrated momentum scale is a key issue. A large
  sample of $J/\psi$ dimuon decays was used to calibrate energy loss and magnetic
  field (Fig.~\ref{masscalib}). Slopes in the transverse momentum dependence of
  the $J/\psi$ mass are attributed to uncorrected energy loss. The corrections are
  adjusted to account for material missing in the description of the detector. The
  overall mass shift with respect to the well measured world average $J/\psi$ mass
  is used to fine tune the magnetic field. The procedure was validated using
  different final states (e.g.~$D^0 \rightarrow K\pi$ decays shown in
  Fig.~\ref{masscalib}).

\begin{figure}[tb]
   \setlength{\unitlength}{1cm}
   \begin{picture}(15.0,3.9)(0.0,0.0)
   \put(0.4,-0.8){\psfig{file=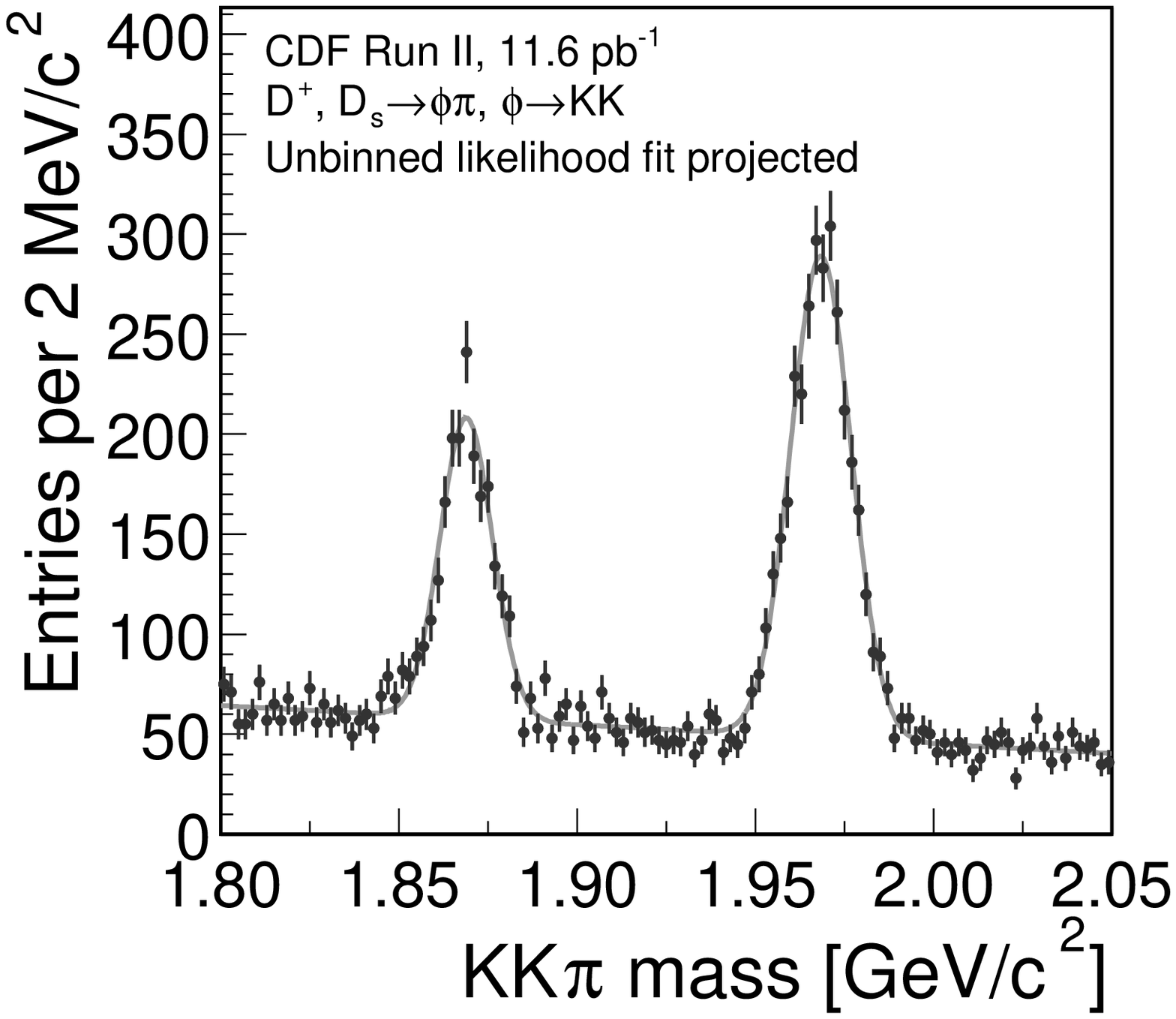,width=5.5cm}}
   \put(6.9,-0.7){\psfig{file=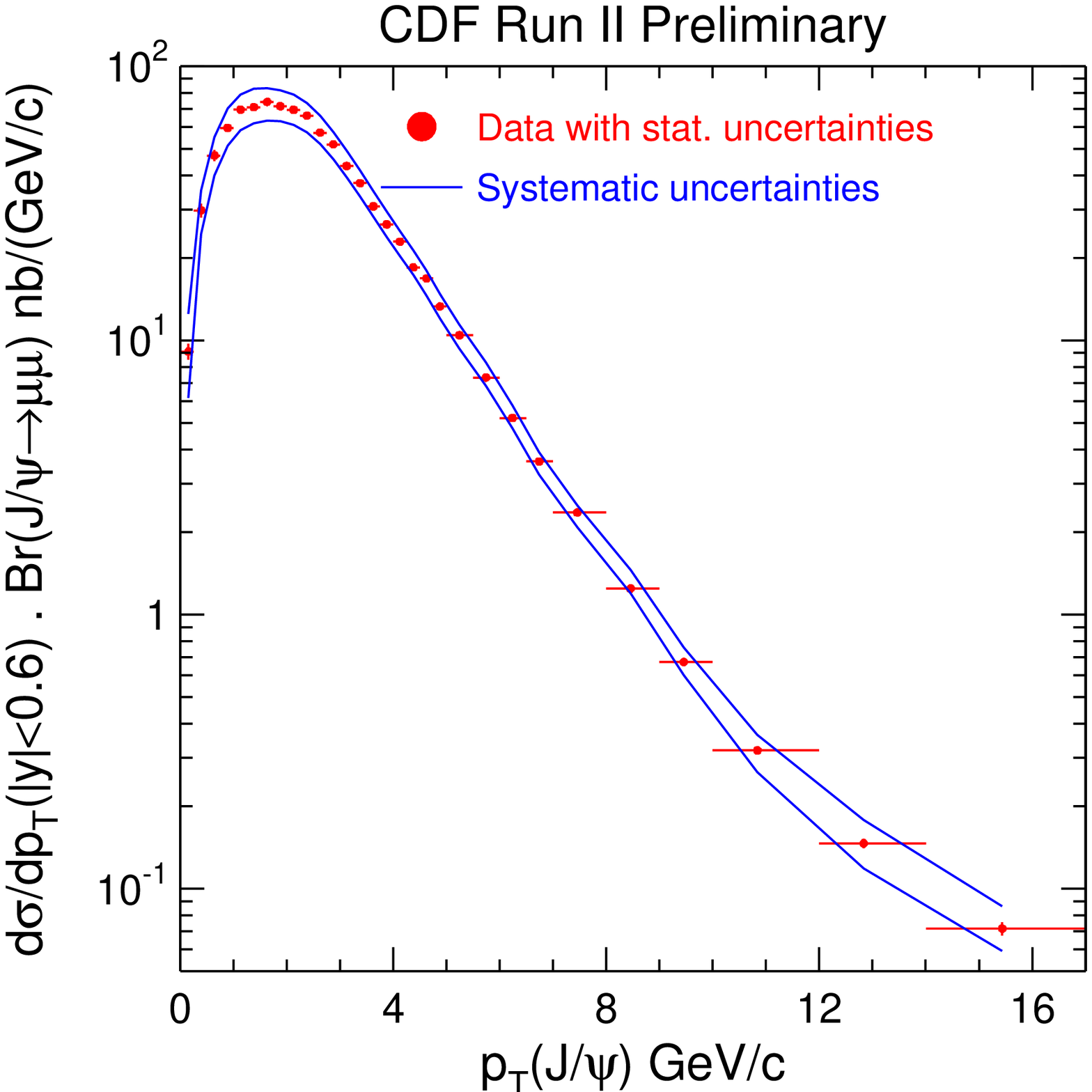,width=5.3cm}}
   \end{picture}
   \vspace*{8pt}
   \caption{{\em Left:} The reconstructed $D^+ \rightarrow \phi\pi^+$ and $D_s^+ \rightarrow \phi\pi^+$ mass
            distribution. {\em Right:} The inclusive $J/\psi$ cross section
            differential in $p_T$ for $|y(J/\psi)|<0.6$.}
   \label{dsignals}
  \end{figure}

  The mass difference measurement relies on $D_s^+$ and $D^+$ decays into
  $\phi\pi^+$ with $\phi\rightarrow K^+K^-$, as shown in Fig.~\ref{dsignals}. Using
  the same decay mode has the advantage of cancelling systematic uncertainties. The
  unbinned likelihood fit of the mass spectrum results in $m_{D_s^+}-m_{D^+} =
  99.41 \pm 0.38 \mbox{(stat)} \pm 0.21 \mbox{(sys)}\:\mbox{MeV}$, with
  uncertainties comparable to the world average\cite{pdg}. This measurement
  constitutes the first published Run~II result from CDF\cite{mass}.

\subsubsection{Cabibbo Suppressed Decays and CP Violation}

\begin{figure}[tb]
   \setlength{\unitlength}{1cm}
   \begin{picture}(15.0,2.8)(0.0,0.0)
   \put(-0.1,-0.8){\psfig{file=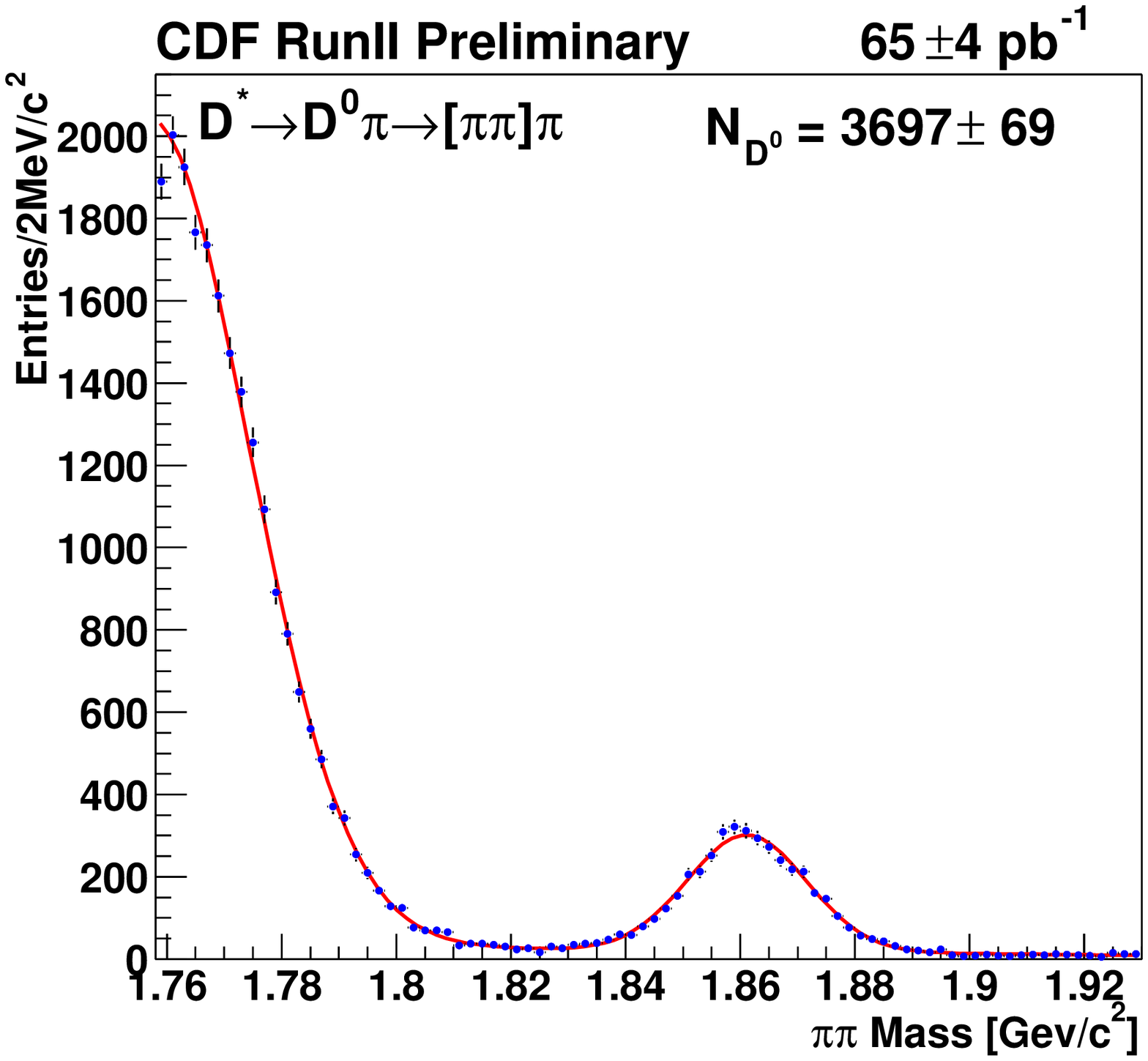,width=4.5cm}}
   \put(4.2,-0.8){\psfig{file=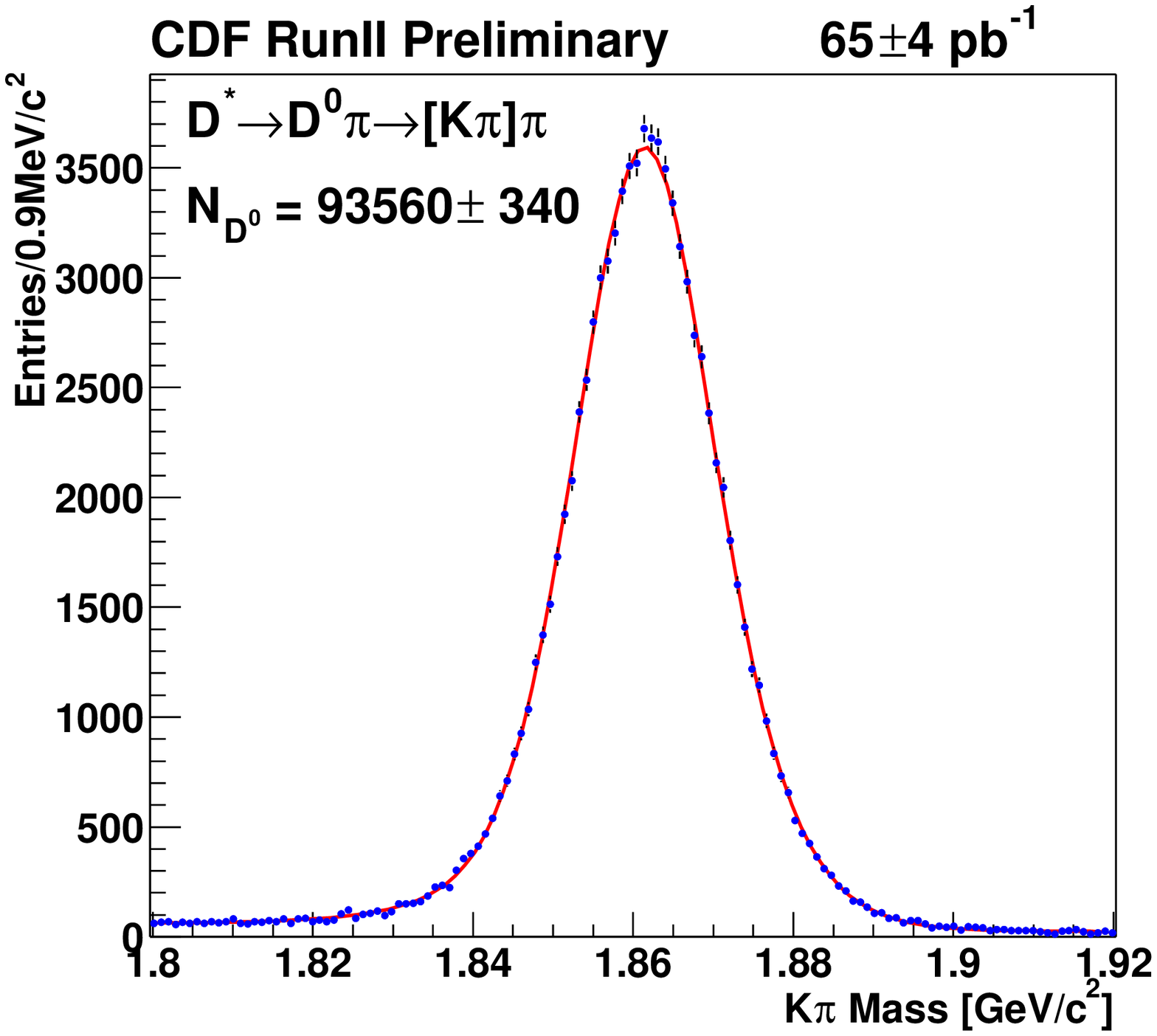,width=4.5cm}}
   \put(8.5,-0.8){\psfig{file=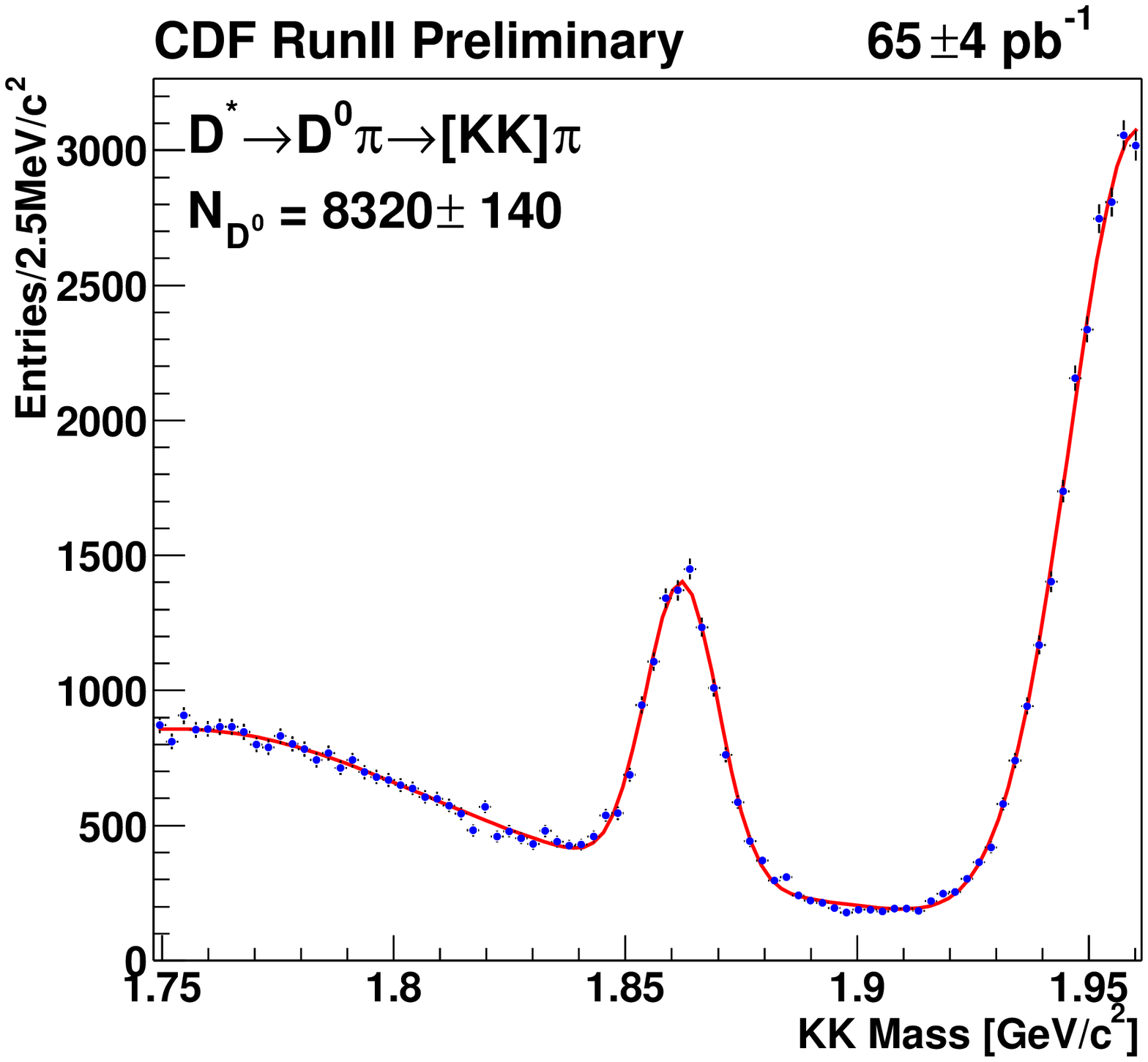,width=4.5cm}}
   \end{picture}
   \vspace*{8pt}
   \caption{Reconstructed $D^0$ decay modes for branching ratio and CP asymmetry
            measurements. Note that in the $KK$ (right) and $\pi\pi$ (left) modes
            the reflections are well separated.}
   \label{cabibbo}
  \end{figure}

  Utilizing the huge sample of $D^0$ mesons in $65\:\mbox{pb}^{-1}$ integrated
  luminosity collected with the secondary vertex trigger, relative branching
  fractions are measured,
  $\frac{\Gamma(D^0\rightarrow K^+K^-)}{\Gamma(D^0\rightarrow K^+\pi^-)} =
         9.38 \pm 0.18 \mbox{(stat)}\pm 0.10 \mbox{(sys)}$\%
  and
  $ \frac{\Gamma(D^0\rightarrow \pi^+\pi^-)}{\Gamma(D^0\rightarrow K^+\pi^-)} =
        3.686 \pm 0.076 \mbox{(stat)}\pm 0.036 \mbox{(sys)}$\%,
  comparing favorably with the current best measurement\cite{focus}. In the
  analysis, the $D^0$ candidate is combined with a charged slow pion to form
  a $D^\ast$ meson; in this way, backgrounds are reduced, and the charge of
  the slow pion from the $D^\ast$ decay serves as an unbiased tag of the
  $D^0$ flavor. Examples of the reconstructed decays are shown in
  Fig.~\ref{cabibbo}.

  The CP violating decay rate asymmetries
  $A=\frac{\Gamma(D^0\rightarrow f) - \Gamma(\bar{D^0}\rightarrow f)}{
           \Gamma(D^0\rightarrow f) + \Gamma(\bar{D^0}\rightarrow f)}$
  are also measured. It is found that
     $A(D^0\rightarrow K^+K^-) = 2.0 \pm 1.7 \mbox{(stat)}\pm 0.6\mbox{(sys)}$\% and
     $A(D^0\rightarrow \pi^+\pi^-) = 3.0 \pm 1.9 \mbox{(stat)}\pm 0.6\mbox{(sys)}$\%,
  comparable to previous measurements\cite{cleo}.

\subsubsection{Search for the FCNC Decay $D^0 \rightarrow \mu^+\mu^-$}

  The search for the flavor changing neutral current (FCNC) decay $D^0
  \rightarrow \mu^+\mu^-$ is another example of an analysis that greatly benefits
  from the SVT trigger, by providing the well measured normalization mode $D^0
  \rightarrow \pi^+\pi^-$. The branching ratio is ${\cal O}(10^{-13})$ in the
  Standard Model, but can be enhanced up to ${\cal B}(D^0 \rightarrow \mu^+\mu^-)
  \simeq 3.5\cdot 10^{-6}$ in R-parity violating models of Supersymmetry.

\begin{figure}[tb]
   \setlength{\unitlength}{1cm}
   \begin{picture}(15.0,3.8)(0.0,0.0)
   \put(0.8,-0.7){\psfig{file=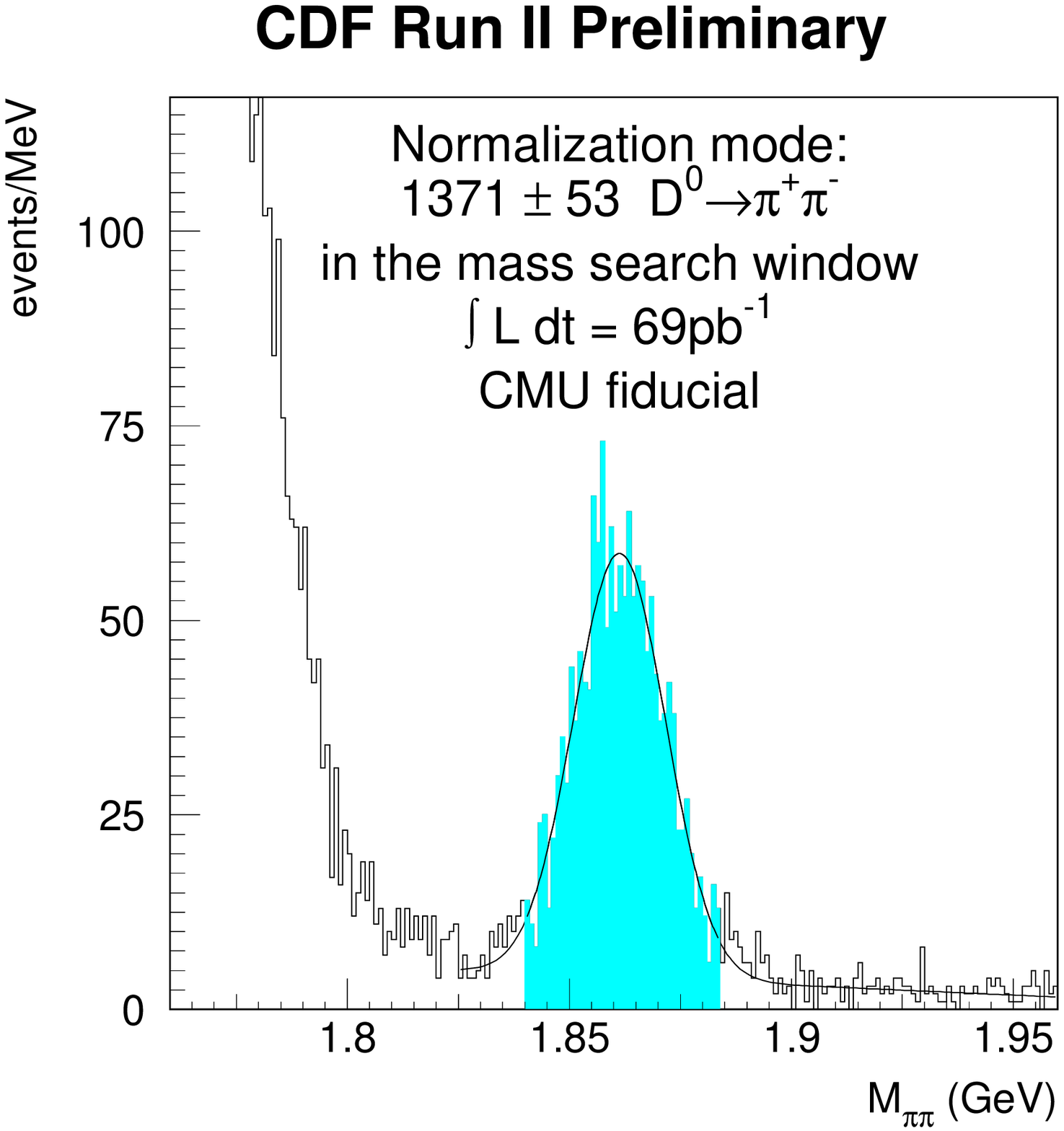,width=5.1cm}}
   \put(6.6,-0.7){\psfig{file=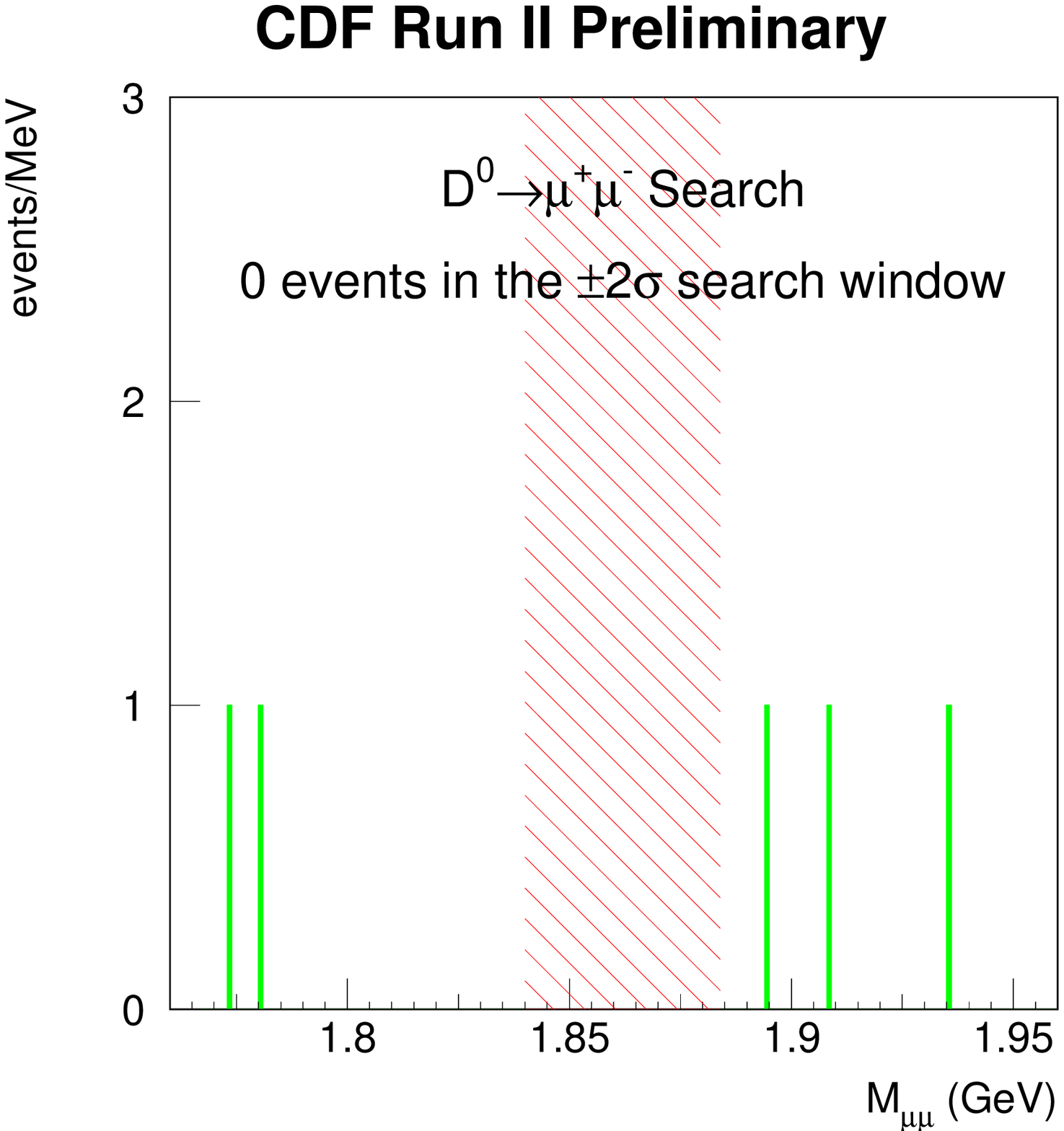,width=5.1cm}}
   \end{picture}
   \vspace*{8pt}
   \caption{Search for $D^0 \rightarrow \mu^+\mu^-$. {\em Left:} Mass spectrum for the
            normalization mode $D^0 \rightarrow \pi^+\pi^-$; the search window is
            indicated by the shaded area. {\em Right:} Dimuon candidate events; the
            search window is indicated by the hatched area.}
   \label{dmumu}
  \end{figure}

  In a data sample corresponding to an integrated luminosity of
  $69\:\mbox{pb}^{-1}$ no candidate event is observed (Fig.~\ref{dmumu}) with
  $1.7\pm 0.7$ background events expected. After correcting for relative
  acceptance an upper limit of $ {\cal B}(D^0 \rightarrow \mu^+\mu^-) \leq
  2.4\cdot 10^{-6}$ at 90\% CL is set, improving the current best
  limit\cite{mumu} of $4.1\cdot 10^{-6}$.

\subsubsection{Measurement of the $J/\psi$ Cross Section}

  One of the surprises of Run~I was the direct production cross section for
  $J/\psi$ and $\psi(2S)$ mesons\cite{psirun1}, which turned out to be orders
  of magnitude larger than the theoretical expectation in the Color Singlet
  Model. While later calculations within the framework of non-relativistic QCD,
  including intermediate color octet states, are in broad agreement with the
  data, there is continued interest in the subject; in particular measurements
  of the $J/\psi$ and $\psi(2S)$ polarization appear to be in conflict with the
  theory, albeit not with convincing statistical significance\cite{psipol}.

  For Run~II the muon trigger momentum thresholds at CDF were lowered to
  \mbox{$\geq 1.4\:\mbox{GeV}$}, thus allowing to trigger on $J/\psi$'s at rest for the
  first time. Using a data sample of $39.7\:\mbox{pb}^{-1}$ the inclusive
  differential cross section in bins of the $J/\psi$ transverse momentum has
  been measured for $J/\psi$ rapidities $|y(J/\psi)|<0.6$
  (Fig.~\ref{dsignals}). The region $p_T < 5\:\mbox{GeV}$ is covered for the
  first time. The total cross section has been determined to be
  $\sigma(p\bar{p} \rightarrow J/\psi X, |y(J/\psi)|<0.6) \cdot {\cal B}(J/\psi
  \rightarrow \mu^+\mu^-) = 240 \pm 1 \mbox{(stat)} ^{+35}_{-28} \mbox{(sys)}
  \:\mbox{nb}$.

  Future measurements will include the determination of the prompt production
  cross section; the inclusive cross section includes $J/\psi$'s from $b$ decays
  amounting to anything between 0\% at small $p_T$ and 50\% at large $p_T$. The
  prompt $J/\psi$ cross section has contributions from $\chi_c$ and $\psi(2S)$ as
  well as the direct component, which has been measured\cite{chi} to be $64 \pm
  6$\% of the prompt cross section in Run~I. In addition, more precise data at
  large $p_T$, polarization measurements, and similar measurements of $\psi(2S)$,
  $\Upsilon(1S)$, $\Upsilon(2S)$, and $\Upsilon(3S)$ production will shed further
  light on the production mechanisms for heavy vector mesons.

\subsection{Beauty}

  The major advantage of hadron colliders in the area of $b$ physics is the huge
  production cross section: $\sigma_{b\bar{b}}\simeq 100\:\mu\mbox{b}$ at
  $\sqrt{s} = 2\:\mbox{TeV}$, compared to $\simeq 1\:\mbox{nb}$ for $e^+e^-$
  colliders running on the $\Upsilon(4S)$ resonance. Making use of this advantage
  requires a very sophisticated trigger system, which is available in Run~II.
  Another benefit of experiments at hadron colliders is the accessibility of all
  species of $b$ hadrons --- hadron colliders are a unique laboratory for the
  study of the $B_s$, $B_c$, and $\Lambda_b$. An extensive review of the Run~II
  prospects can be found in Ref.~\refcite{bws}.

  Although most of the important measurements related to the
  Cabibbo-Kobayashi-Maskawa matrix, CP violation, and $B_s$ mixing require
  significantly larger amounts of data than available to date, a number of
  competitive benchmark measurements have already been performed.

  \subsubsection{\mbox{Measurement of $B$ Meson Masses and Lifetimes in Exclusive $J/\psi$ Modes}}

\begin{figure}[tb]
   \setlength{\unitlength}{1cm}
   \begin{picture}(15.0,3.3)(0.0,0.0)
   \put(0.0,-0.7){\psfig{file=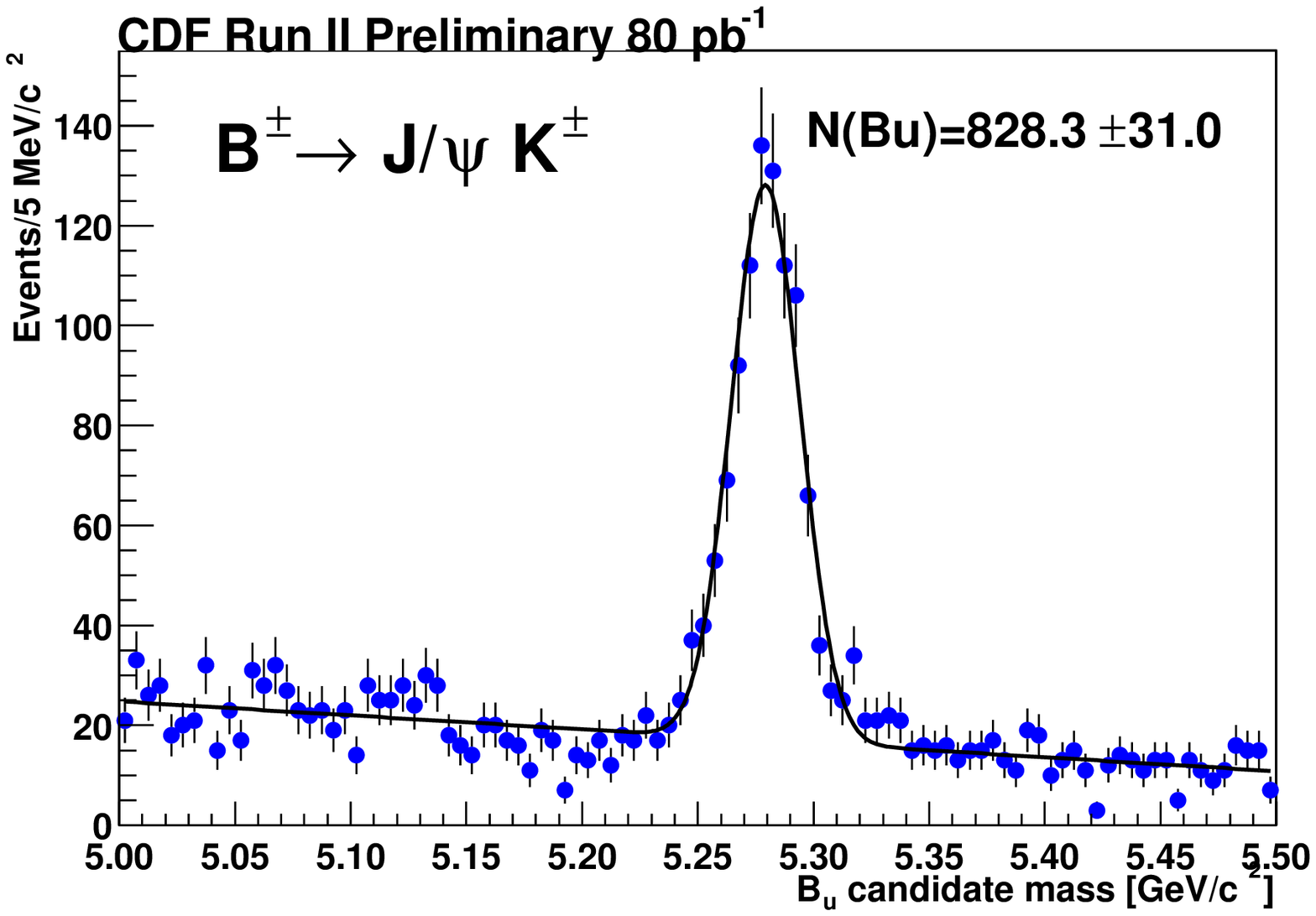,width=6.1cm}}
   \put(6.6,-0.9){\psfig{file=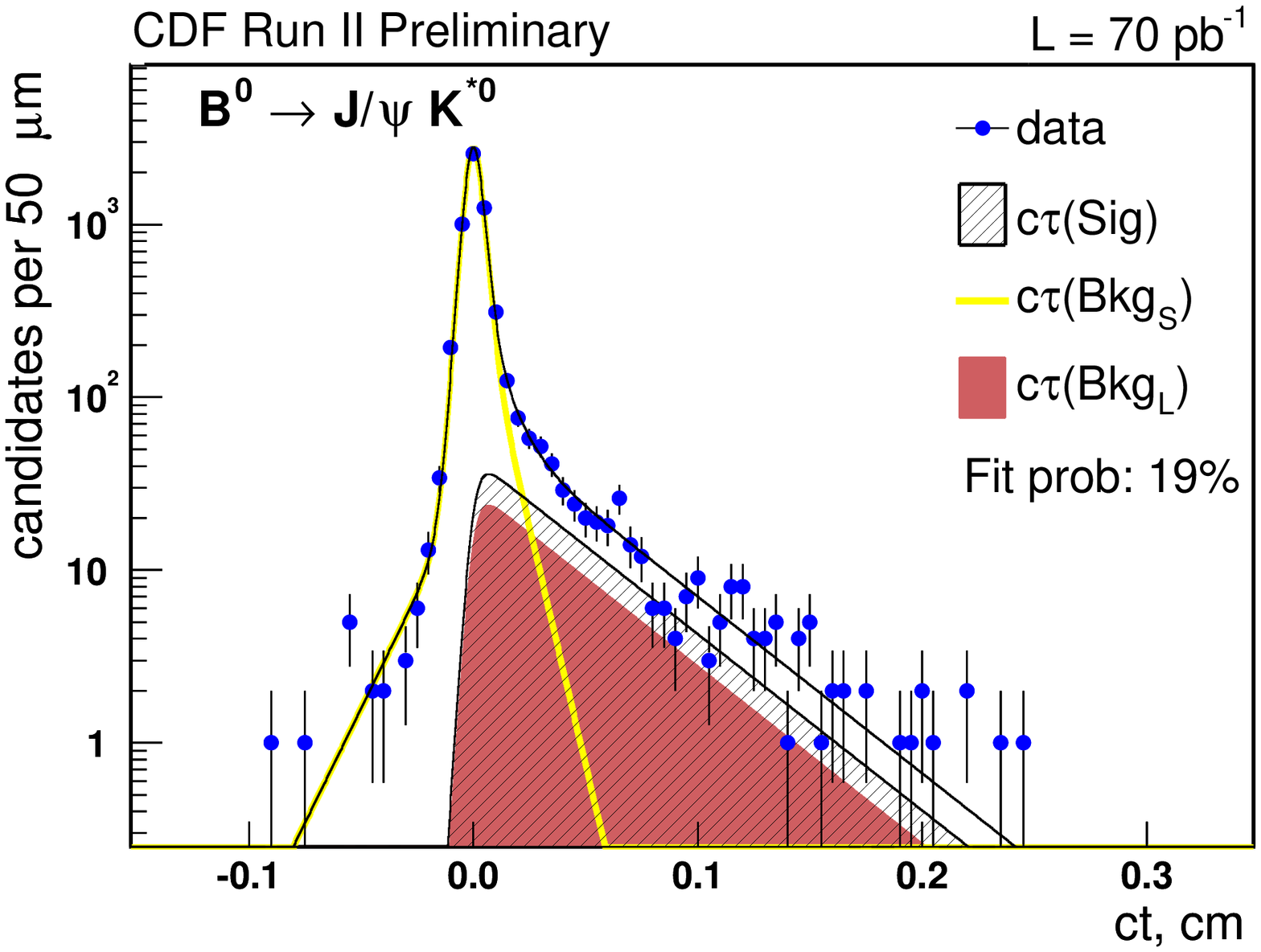,width=6.0cm}}
   \end{picture}
   \vspace*{8pt}
   \caption{{\em Left:} $B^+$ signal reconstructed in the $J/\psi K^+$ channel.
            {\em Right:} The proper decay length distribution of $B^0$ candidate events. The
             result of the maximum likelihood fit is overlaid.}
   \label{bplots}
  \end{figure}

  The masses of the groundstate $B$ mesons have been determined in the exclusive
  $J/\psi$ decay modes, $B^+ \rightarrow J/\psi K^+$ (Fig.~\ref{bplots}),
  $B^0 \rightarrow J/\psi K^{\ast 0}$, and
  $B_s^0 \rightarrow J/\psi \phi$, using a data sample corresponding to $80\:\mbox{pb}^{-1}$.
  The results for the $B^+$ and $B^0$,
  $m(B^+) = 5279.32 \pm 0.68 \mbox{(stat)} \pm 0.94 \mbox{(sys)} \:\mbox{MeV}$ and
  $m(B^0) = 5280.30 \pm 0.92 \pm 0.96\:\mbox{MeV}$
  compare favorably with present world averages\cite{pdg}:
  $m(B^+) = 5279.0 \pm 0.5 \:\mbox{MeV}$ and
  $m(B^0) = 5279.4 \pm 0.5 \:\mbox{MeV}$.
  The $B_s^0$ measurement,
  $m(B_s^0) = 5265.50 \pm 1.29 \pm 0.94\:\mbox{MeV}$,
  constitutes the world best measurement (PDG\cite{pdg}: $m(B_s^0) = 5269.6 \pm 2.4 \:\mbox{MeV}$).

  The same decay modes have been used to measure $B$ meson lifetimes. From an unbinned
  likelihood fit to the proper decay length the following lifetimes are extracted:
  $c\tau_{B^+} = 470 \pm 20 \pm 6\:\mu\mbox{m}$,
  $c\tau_{B^0} = 425 \pm 28 \pm 6\:\mu\mbox{m}$, and
  $c\tau_{B_s} = 379 \pm 59 \pm 6\:\mu\mbox{m}$.
  The proper decay length distribution for the $B^0$ sample is shown in
  Fig.~\ref{bplots}.

  \subsubsection{Hadronic Decays}

  A large number of additional exclusive $b$ hadron decays have been identified in
  the early Run~II data, thanks to the flexible trigger scheme of the upgraded CDF
  detector, including for example the purely hadronic decays $B \rightarrow
  K\pi$/$\pi\pi$/$KK$, $B_s \rightarrow D_s^{(\ast)} \pi$ with $D_s \rightarrow
  \phi\pi$ and $\phi\rightarrow KK$, $B^\pm \rightarrow \phi K^\pm$, and
  $\Lambda_b \rightarrow \Lambda_c^+\pi^-$ with $\Lambda_c^+\pi^- \rightarrow p
  K^- \pi^+$, and will be used for a rich $b$ physics program. The observed signal
  for $\Lambda_b \rightarrow \Lambda_c^+\pi^-$ is shown in Fig.~\ref{lbwe}. At the
  time of writing, this is the largest sample of fully reconstructed $\Lambda_b$'s
  in existance. In the future, beautiful baryons may turn out to be an excellent
  laboratory for testing the Heavy Quark Effective Theory.

\section{Electroweak and Top Physics}

  \begin{figure}[tb]
   \setlength{\unitlength}{1cm}
   \begin{picture}(15.0,3.8)(0.0,0.0)
   \put(0.8,-0.7){\psfig{file=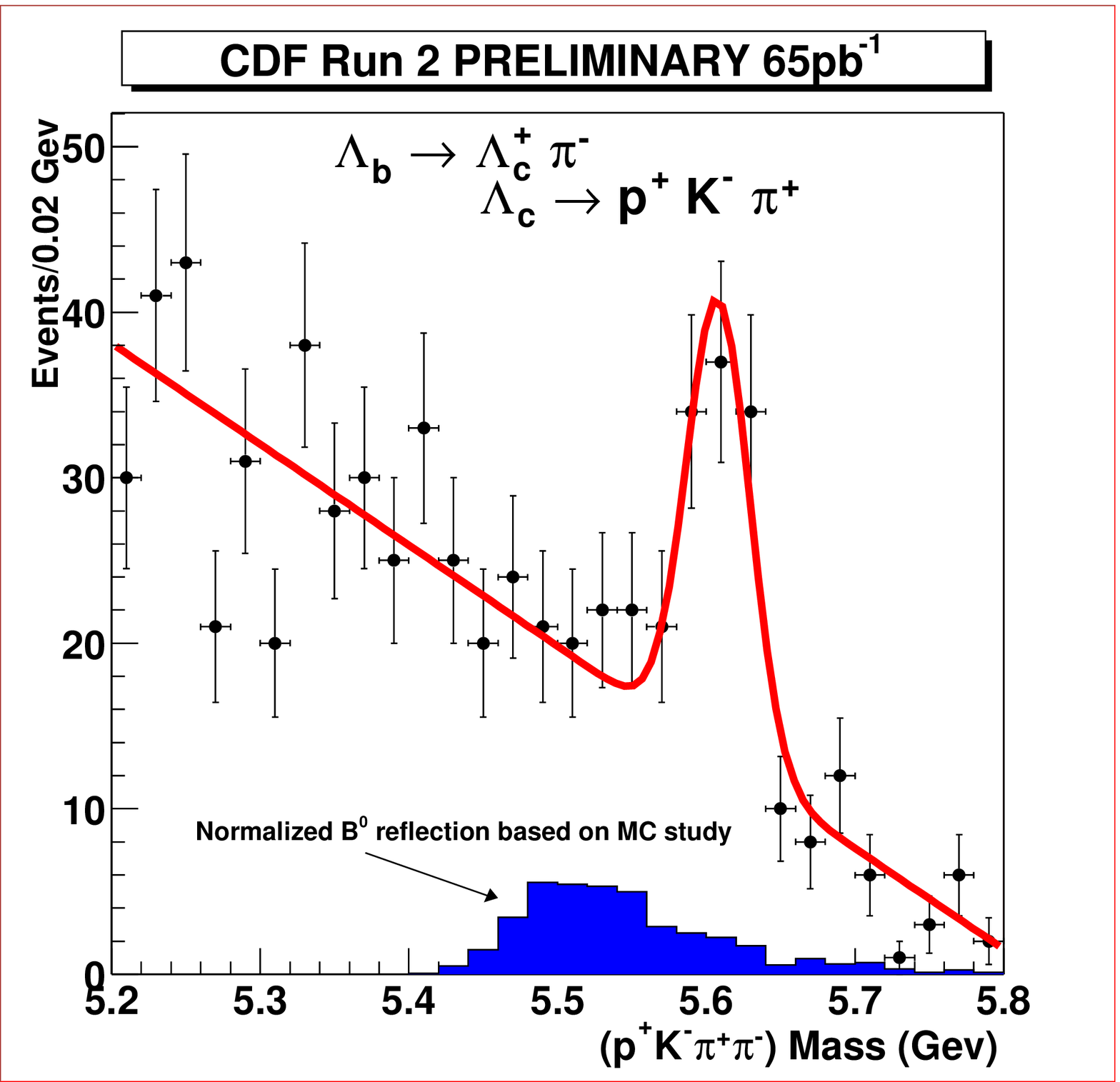,width=4.7cm}}
   \put(6.8,-0.9){\psfig{file=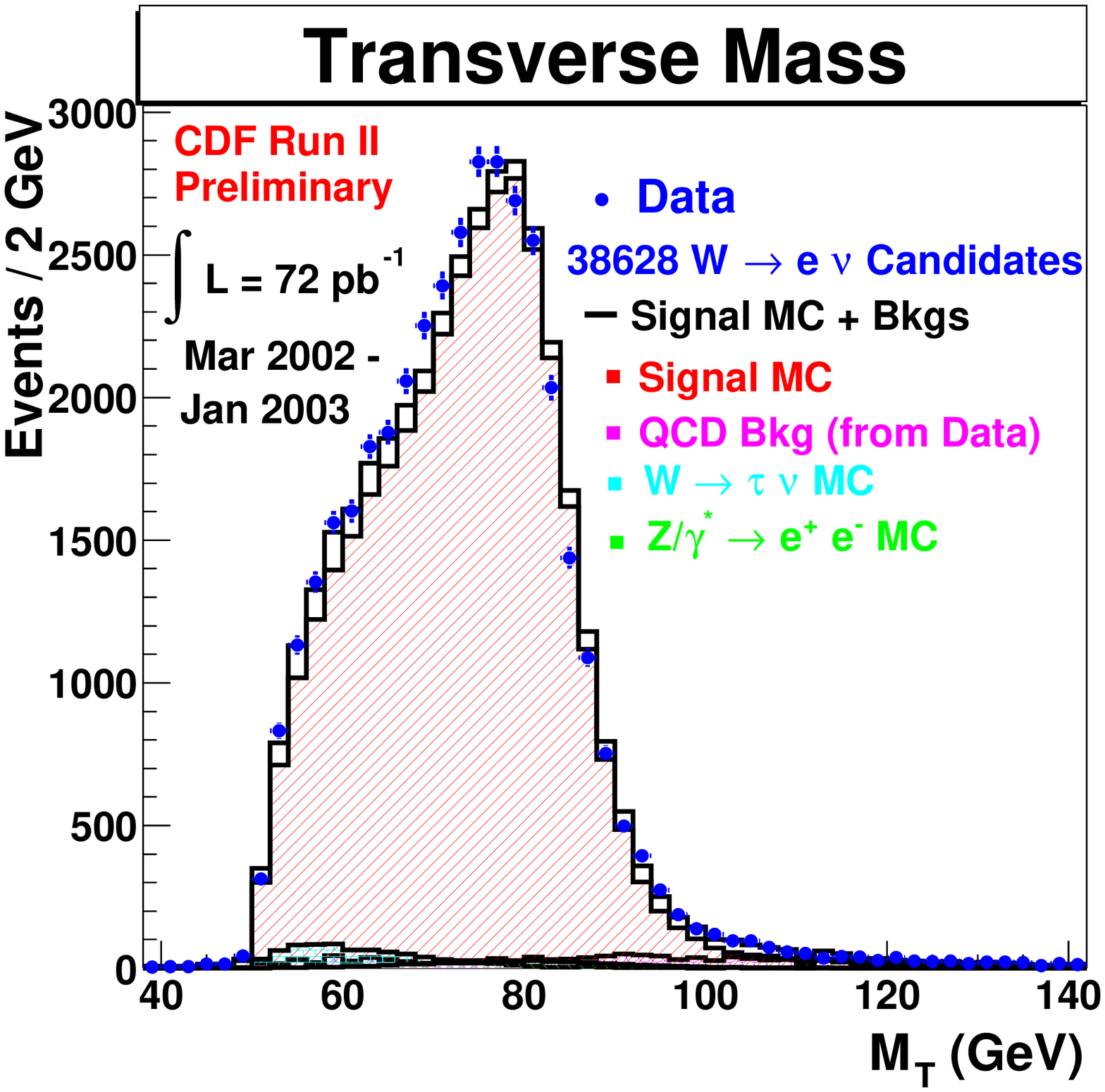,width=5.0cm}}
   \end{picture}
   \vspace*{8pt}
   \caption{{\em Left:} Observation of $\Lambda_b \rightarrow \Lambda_c^+\pi^-$ signal.
            {\em Right:} Transverse mass distribution for $W\rightarrow e\nu$ candidate
             events, overlaid with the expected shape for background and signal.}
   \label{lbwe}
  \end{figure}

\begin{figure}[tb]
   \setlength{\unitlength}{1cm}
   \begin{picture}(15.0,3.6)(0.0,0.0)
   \put(0.0,-0.8){\psfig{file=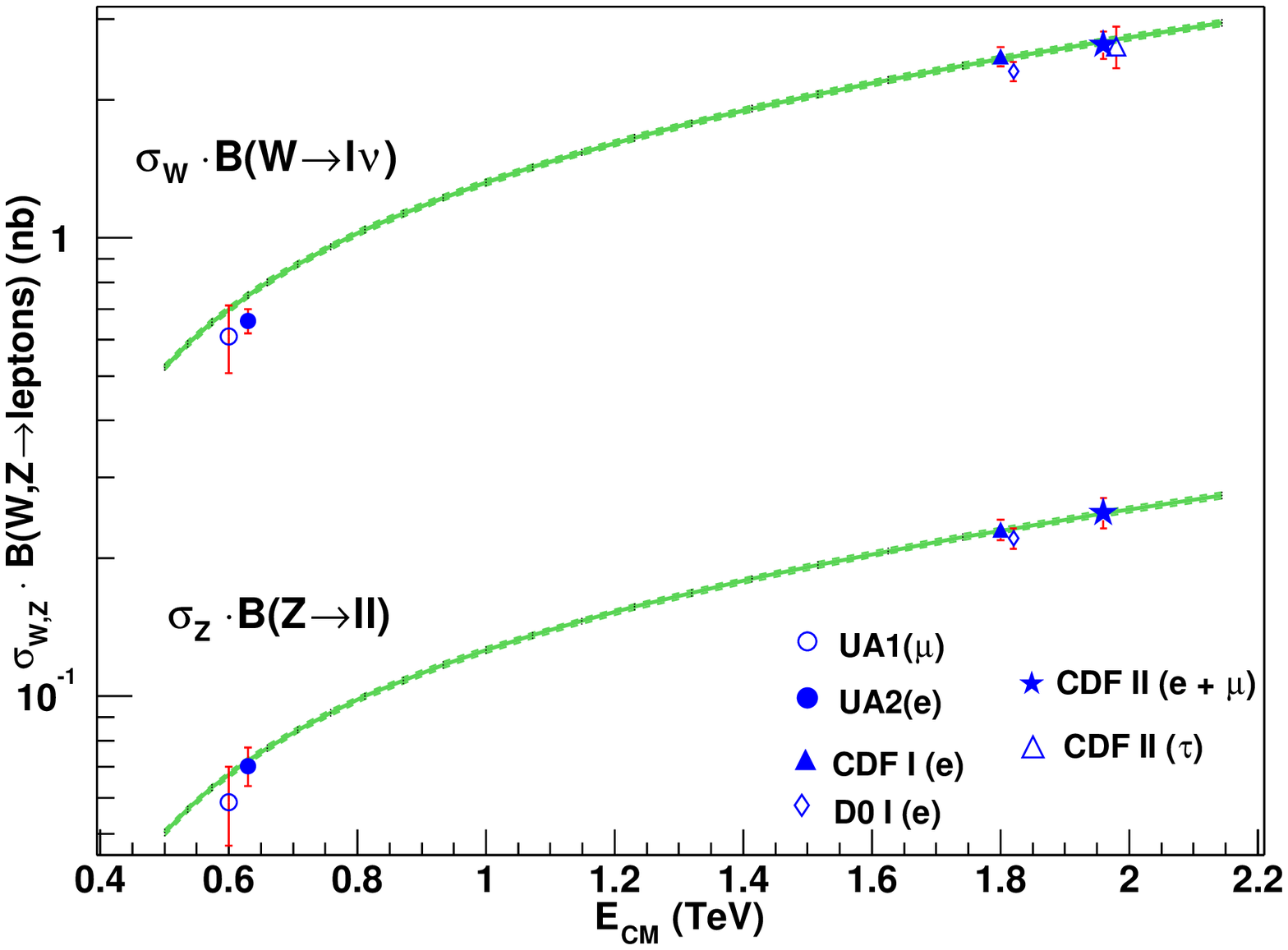,width=6.2cm}}
   \put(6.1,-0.9){\psfig{file=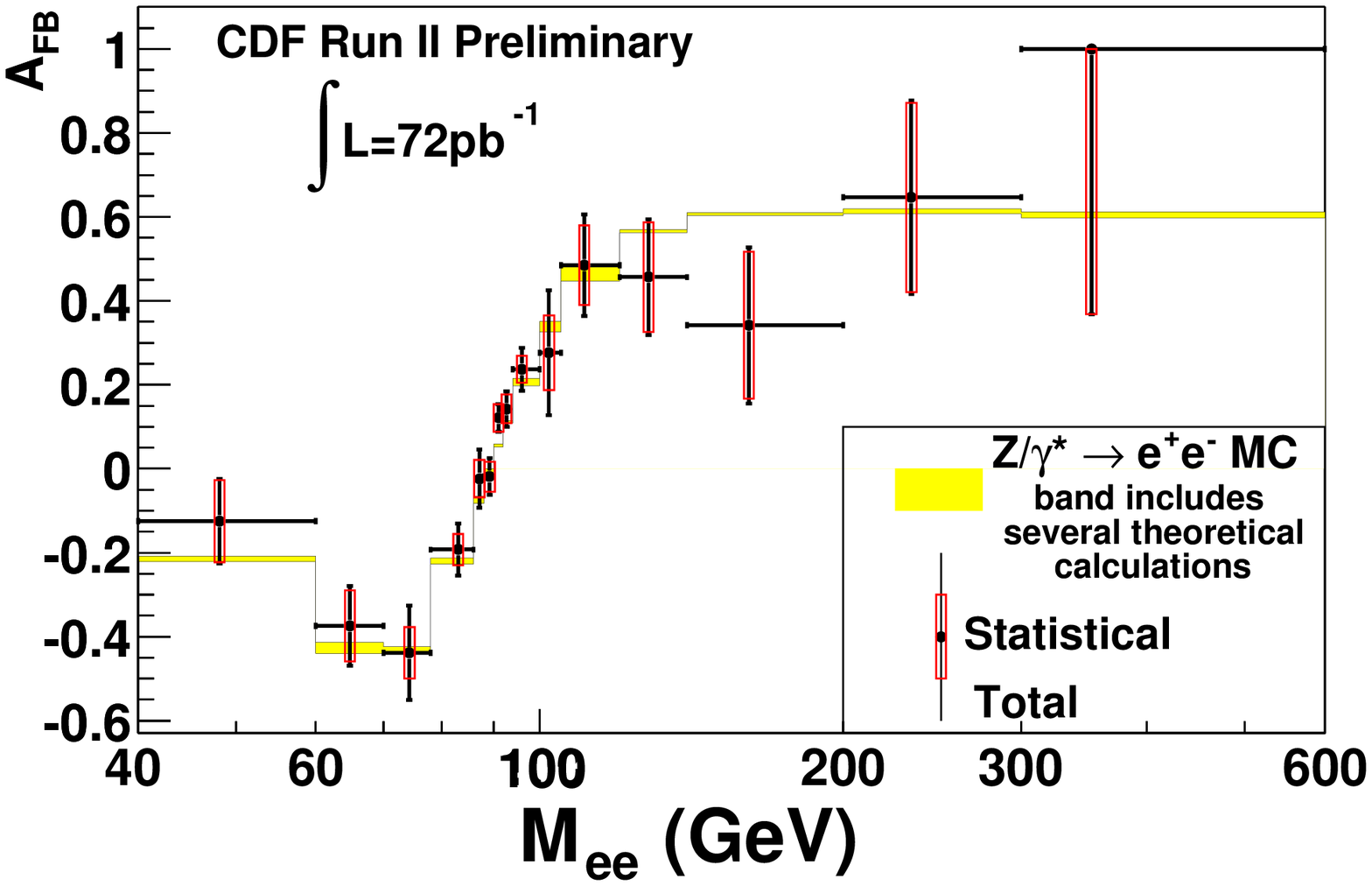,width=6.7cm}}
   \end{picture}
   \vspace*{8pt}
   \caption{{\em Left: Cross section measurements for the $W$ and $Z$ boson together
            with the NNLO prediction}.
            {\em Right:} Measurement of the forward-backward asymmetry $A_{FB}$ of
            electron positron pairs. Predictions by the PYTHIA Monte Carlo
            simulation are shown as a band.}
   \label{wzxsec}
  \end{figure}

\begin{figure}[tb]
   \setlength{\unitlength}{1cm}
   \begin{picture}(15.0,3.8)(0.0,0.0)
   \put(0.2,-0.9){\psfig{file=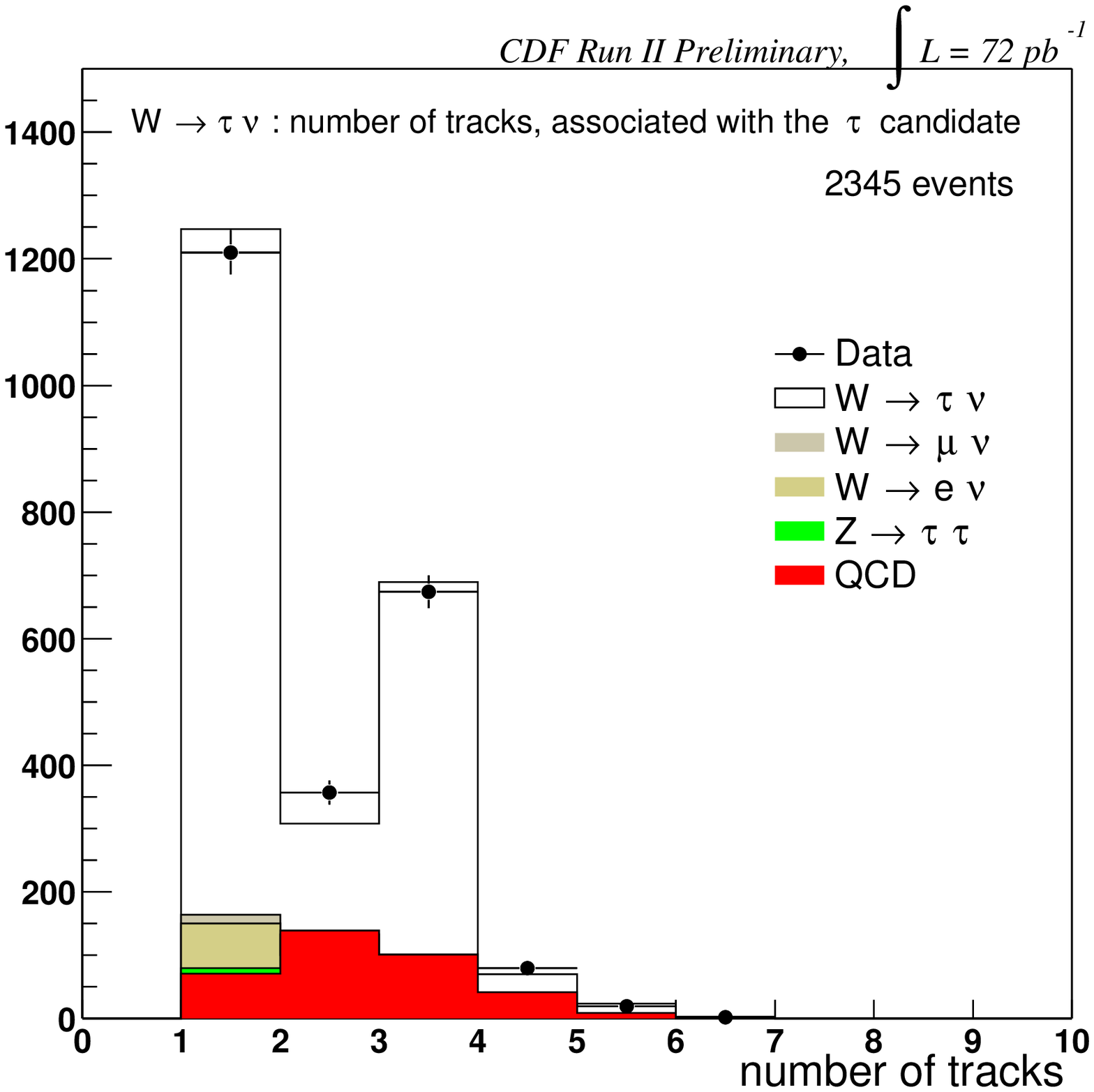,width=5.3cm}}
   \put(6.2,-0.9){\psfig{file=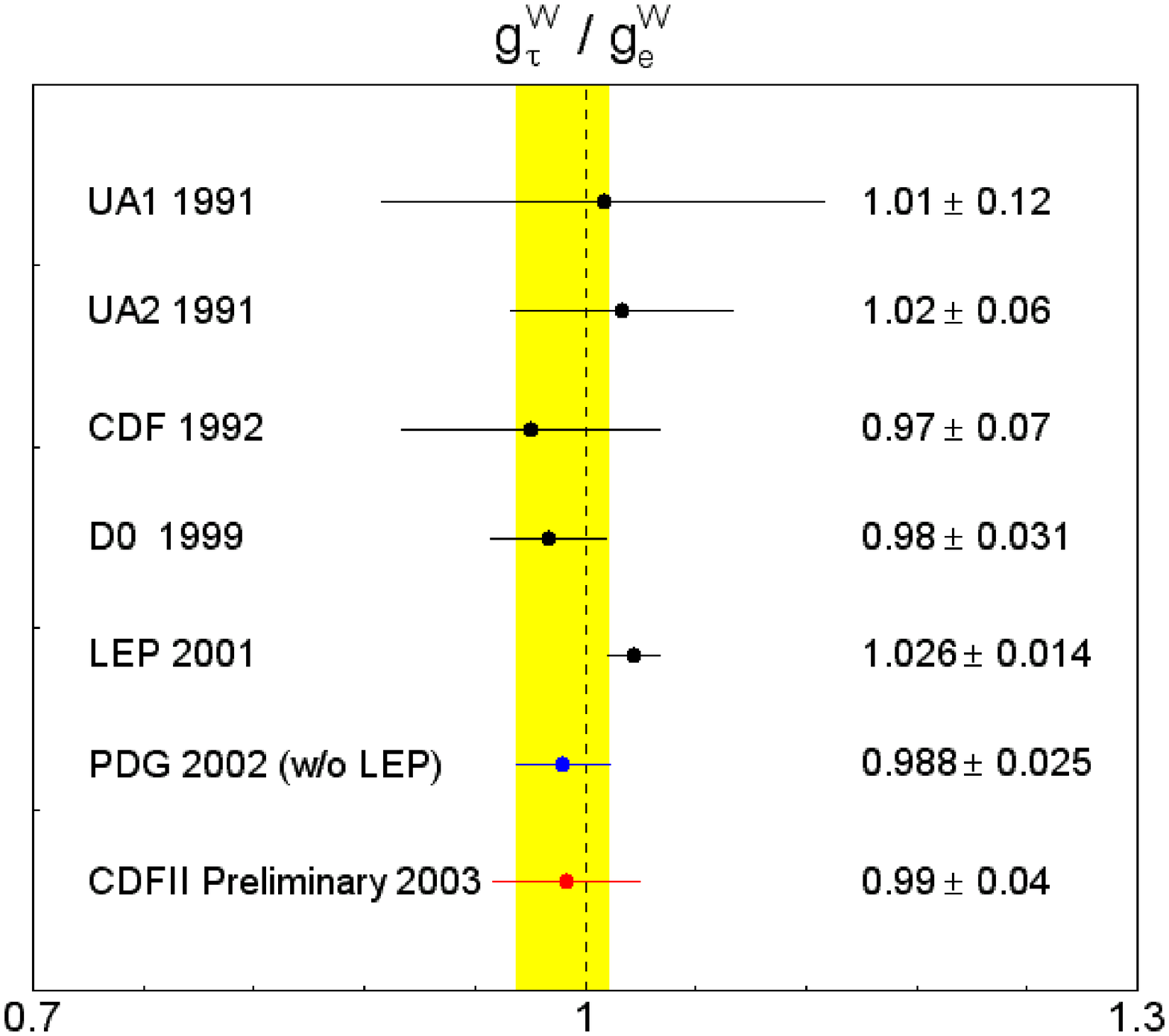,width=5.5cm}}
   \end{picture}
   \vspace*{8pt}
   \caption{{\em Left:} Signal for $W\rightarrow \tau\nu$ seen in the track multitplicity
            distribution for the $\tau$ candidates. {\em Right:}
            The ratio of the couplings of the $W$ to the $\tau$ and $e$ as measured by CDF
            compared to earlier measurements.}
   \label{wtau}
  \end{figure}

  A couple of Run~II benchmark measurements are expected from the electroweak
  sector, e.g.~the precise determination of the $W$ boson mass. $W$'s and $Z$'s
  are collected in large numbers and very efficiently in all possible decay
  channels. Fig.~\ref{lbwe} and \ref{wtau} show examples of $W$ signals in the
  $e$ and $\tau$ channels, and the good understanding of remaining backgrounds.
  The $W$ cross section has been measured in all three lepton channels\cite{ewk} (in the
  electron channel
  $\sigma_W \cdot {\cal B}(W\rightarrow e\nu) = 2.64 \pm 0.01\mbox{(stat)} \pm 0.09\mbox{(sys)}
                                                     \pm 0.16\mbox{(lum)} \:\mbox{nb}$),
  and is in good agreement with next-to-next-to-leading order QCD calculations
  ($2.731 \pm 0.002\:\mbox{nb}$). Good agreement is also found between
  measurements of the $Z$ cross section in the $ee$ and $\mu\mu$ channel and a
  NNLO prediction. A compilation of the cross section measurements is shown in
  Fig.~\ref{wzxsec}.

  The ratio of the couplings of the $W$ to the $\tau$ and $e$ has been measured
  (Fig.\ref{wtau}), which is sensitive to other final states involving $\tau$'s and
  missing transverse energy (charged Higgs decays, production of SUSY particles at
  large $\tan \beta$). CDF's capabilities to collect large samples of high $p_T$
  $\tau$'s are greatly enhanced in Run~II due to the possibility to trigger on
  hadronic $\tau$ decays.

  The reaction $p\bar{p} \rightarrow l^+l^-$, where $l$ stands for an isolated high
  $p_T$ electron or muon, is mediated by virtual photons at low invariant masses
  $M_{l^+l^-}$, by the $Z^0$ around the $Z$ pole, and by $\gamma$-$Z$ interference.
  The presence of both vector and axial-vector couplings of electroweak bosons to
  fermions in the process $q\bar{q} \rightarrow Z^0/\gamma^\ast \rightarrow l^+l^-$
  gives rise to an asymmetry in the polar angle $\theta$ of the electron momentum
  in the center of mass frame of the lepton pair. The forward-backward asymmetry is
  defined as $A_{FB} = (N_F-N_B) / (N_F+N_B)$, where $N_F$ ($N_B$) is the number of
  forward (backward) events with positive (negative) $\cos\theta$. $A_{FB}$ is a
  direct probe of the relative strength of the vector and axial-vector couplings.
  In addition, $A_{FB}$ constrains the properties of any additional heavy neutral
  gauge boson, and is complementary to a direct search via deviations in the
  cross section measurement.
  Using 5438 $e^+e^-$ pairs collected in a data sample corresponding to
  $72\:\mbox{pb}^{-1}$, the forward-backward asymmetry is measured for $40 <
  M_{e^+e^-} < 600\:\mbox{GeV}$. The result is shown in Fig.~\ref{wzxsec}, and
  agrees well with theoretical predictions.

  \paragraph{Top Production}

  The Tevatron experiments are to date the only places where direct measurements of
  the top quark can be performed. At the Tevatron top quarks are produced
  predominantly in pairs through the QCD processes $q\bar{q}\rightarrow t\bar{t}$
  (85\%) and $gg \rightarrow t\bar{t}$ (15\%). Single top quark production via the
  electroweak vertex $Wtb$ is predicted with about half the cross section, but ---
  with final states difficult to extract from the background --- is only expected
  to be observable later in Run~II. With $2\:\mbox{fb}^{-1}$ of integrated
  luminosity, cross section measurements with a precision of 7\% and 20\% are
  expected for $t\bar{t}$ and single top production, respectively.

\begin{figure}[tb]
   \setlength{\unitlength}{1cm}
   \begin{picture}(15.0,3.8)(0.0,0.0)
   \put(0.4,-1.1){\psfig{file=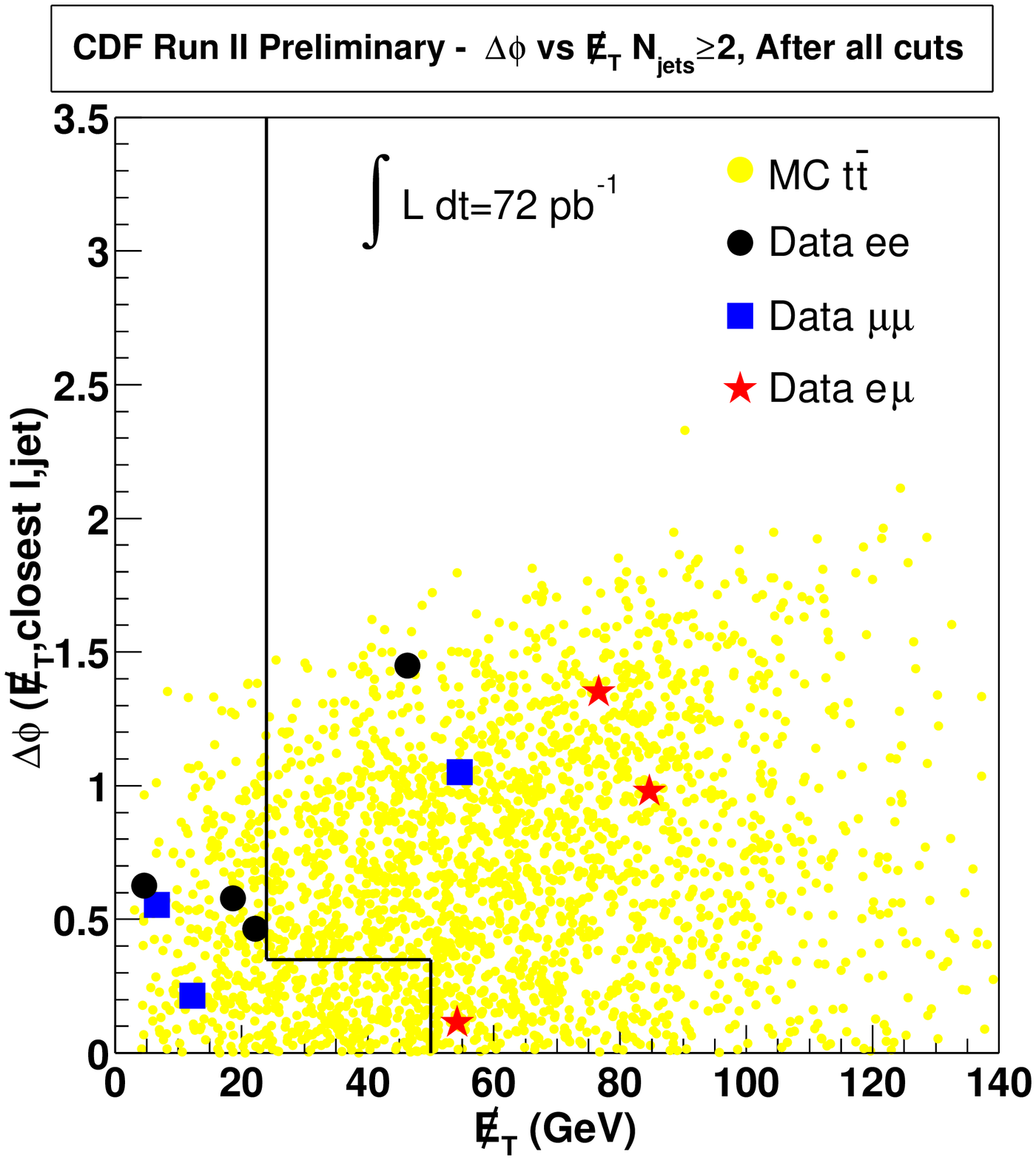,width=5.4cm,height=5.4cm}}
   \put(6.2,-0.6){\psfig{file=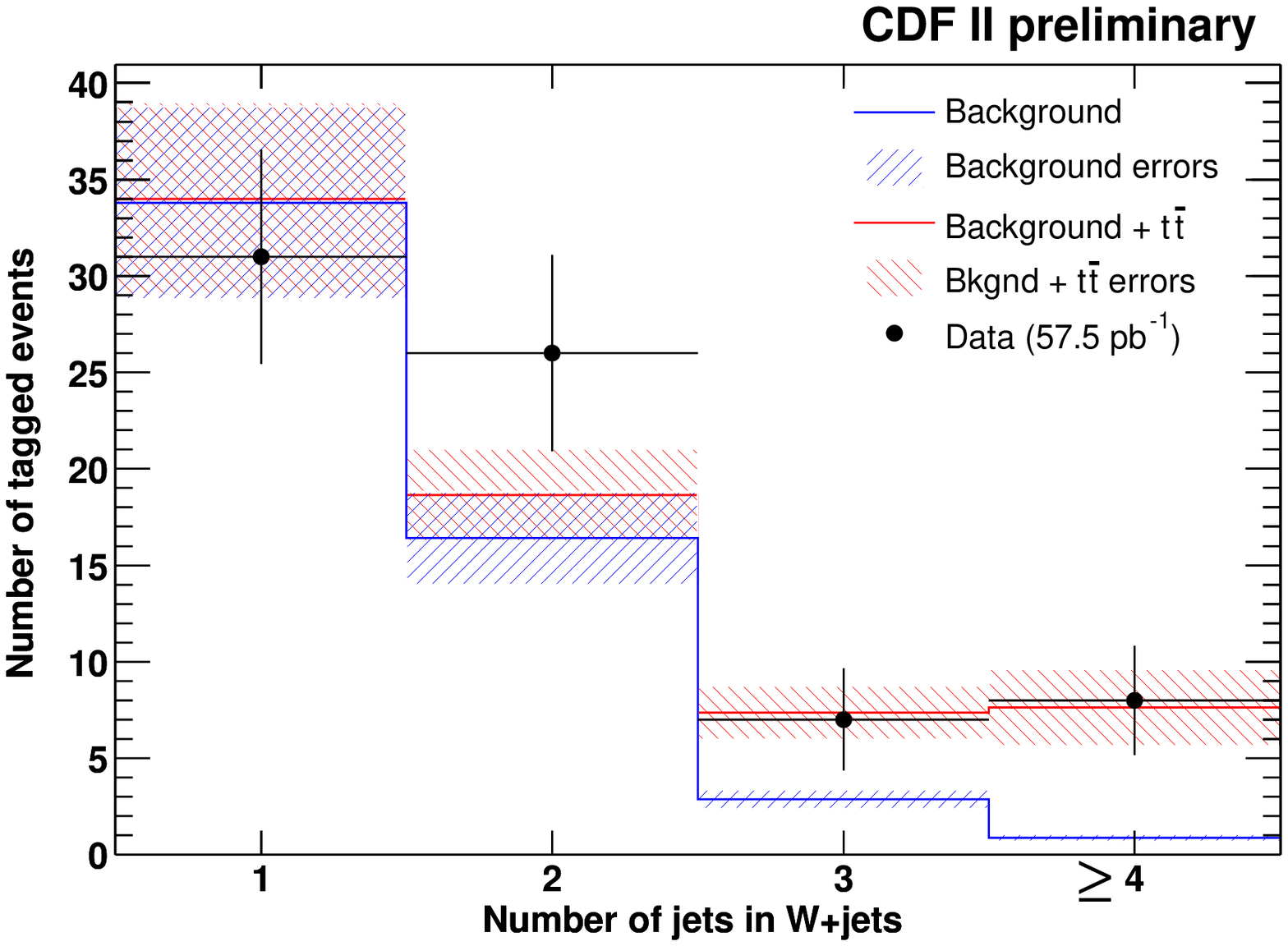,width=6.7cm}}
   \end{picture}
   \vspace*{8pt}
   \caption{{\em Left:} The five $t\bar{t}$ dilepton candidates in the plane
             $\Delta\phi (\not\!\!\!E_T, \mbox{~nearest~} l \mbox{~or~} j)$
             vs.~$\not\!\!\!E_T$ in comparison with the Herwig $t\bar{t}$ Monte
             Carlo simulation.
             {\em Right:} Number of events in the $W +$ jets sample with at least
             one $b$ tag; the 3 and $\geq 4$ bins are used to extract $\sigma_{t\bar{t}}$.}
   \label{top}
  \end{figure}

  Within the Standard Model the top quark decays almost exclusively into $Wb$. In the
  dilepton channel, both $W$'s decay leptonically to $e$ or $\mu$; the branching
  ratio is small (5\%), but so are backgrounds. In the ``lepton $+$ jets'' channel
  one $W$ decays leptonically and the other one hadronically; the branching ratio is
  $\simeq 30$\%. Fig.~\ref{top} shows the candidate events in both channels, for an
  integrated luminosity of $72\:\mbox{pb}^{-1}$ and $58\:\mbox{pb}^{-1}$,
  respectively. The cross section is measured to be
     $\sigma_{t\bar{t}} = 5.3\pm 1.9 \mbox{(stat)}\pm 0.8 \mbox{(sys)}\pm 0.3 \mbox{(lum)} \:\mbox{pb} $
    (lepton $+$ jets) and
     $ \sigma_{t\bar{t}} = 13.2\pm 5.9 \mbox{(stat)}\pm 1.5 \mbox{(sys)}\pm 0.8 \mbox{(lum)} \:\mbox{pb} $
  (dilepton).
  These results are consistent with the next-to-leading order QCD
  prediction\cite{top} of $\sigma_{t\bar{t}} = 6.70^{+0.71}_{-0.88}\:\mbox{pb}$,
  which is about 30\% higher than the corresponding Run~I value due to the higher
  center of mass energy.

\section{QCD}

\begin{figure}[tb]
   \setlength{\unitlength}{1cm}
   \begin{picture}(15.0,3.5)(0.0,0.0)
   \put(-0.2,-0.9){\psfig{file=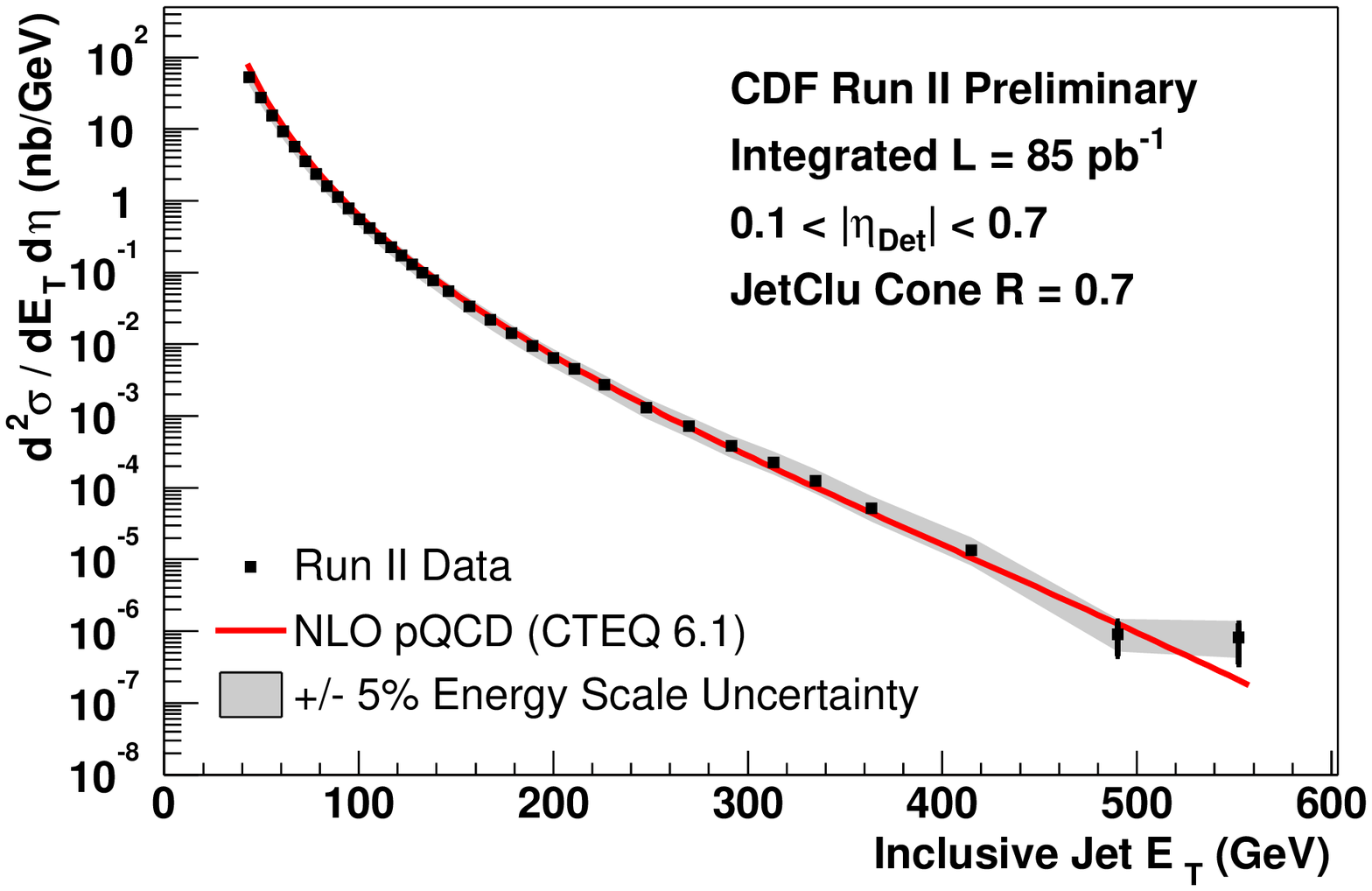,width=6.7cm}}
   \put(6.3,-0.7){\psfig{file=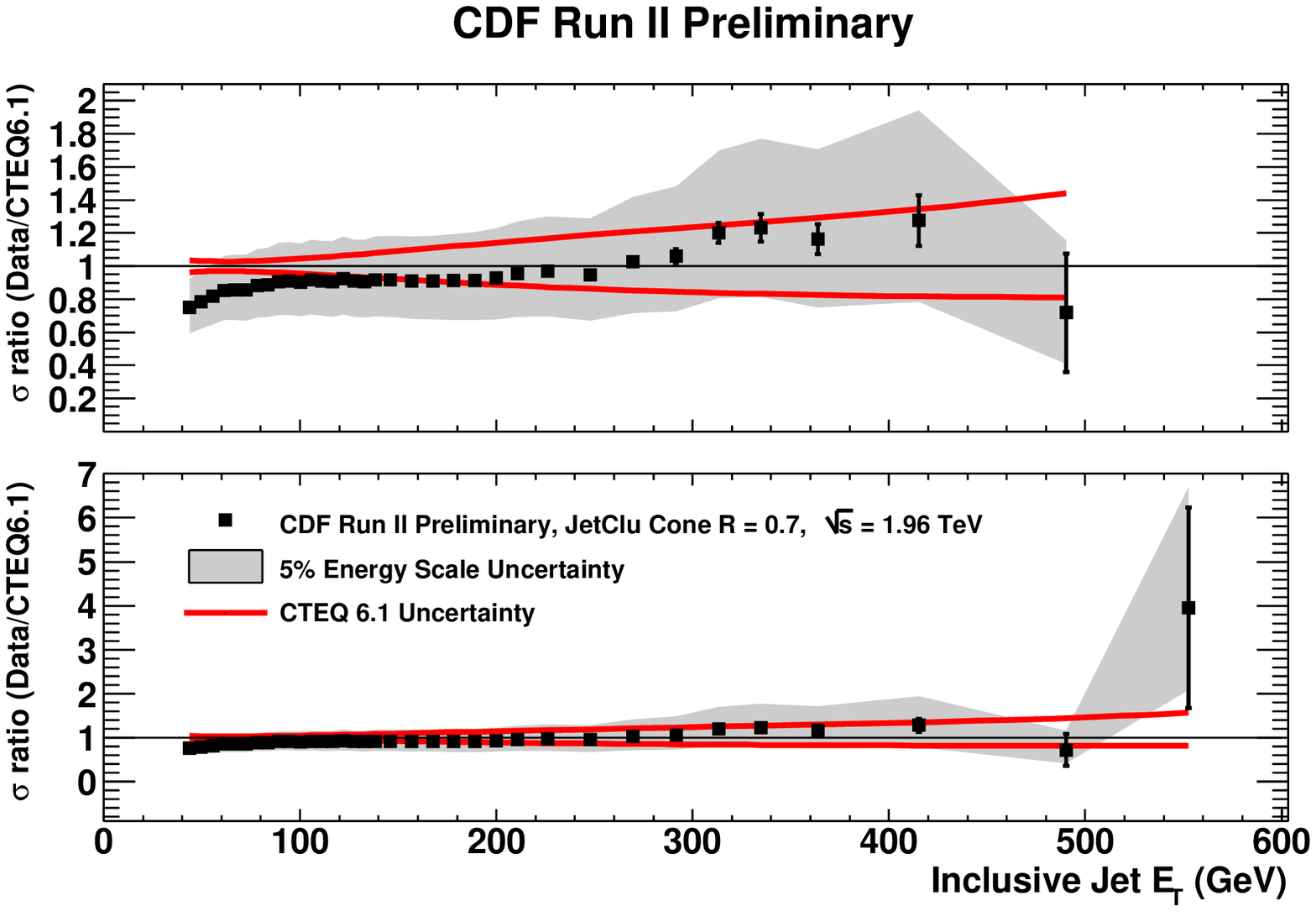,width=6.9cm}}
   \end{picture}
   \vspace*{8pt}
   \caption{{\em Left:} Measurement of the inclusive jet cross section, compared to the NLO
              QCD expectation using the CTEQ 6.1 parton density functions. The
              variation in the cross section due to the 5\% energy scale uncertainty is
              shown as a band. The two lines indicate the uncertainty in the cross section
              prediction due to the PDF. {\em Right:} The ratio of data and theory.}
   \label{jets}
  \end{figure}

  Hadronic jets are one of the key signatures in hadronic collider physics, and
  probe the highest momentum transfer region currently accessible at any
  collider. A first measurement of the inclusive jet $E_T$ cross section in
  Run~II based on an integrated luminosity of $85\:\mbox{pb}^{-1}$ of data has
  been performed. The data span the $E_T$ range from $44$ to $550\:\mbox{GeV}$,
  extending the upper limit by almost $150\:\mbox{GeV}$ compared to Run~I. In
  Fig.~\ref{jets} the measurement is compared to the NLO QCD expectation determined
  using the CTEQ~6.1 parameterization\cite{cteq} of the parton density functions.
  The systematic uncertainty is dominated by the 5\% energy scale uncertainty,
  which is expected to be reduced to of the order of 1\% over the course of
  Run~II.

\begin{figure}[tb]
   \setlength{\unitlength}{1cm}
   \begin{picture}(15.0,4.5)(0.0,0.0)
   \put(-0.1,-0.2){\psfig{file=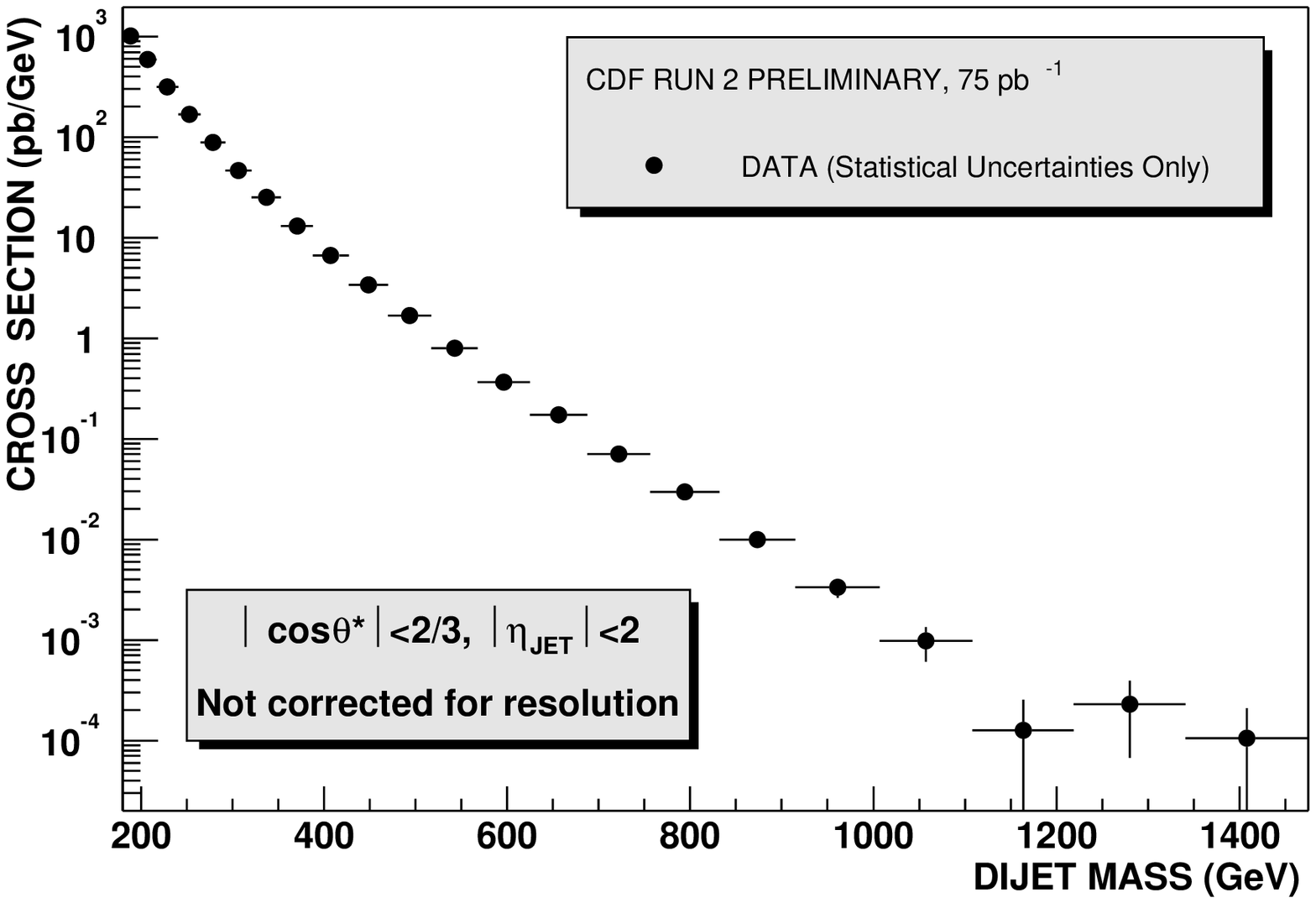,width=6.9cm}}
   \put(6.3,-0.9){\psfig{file=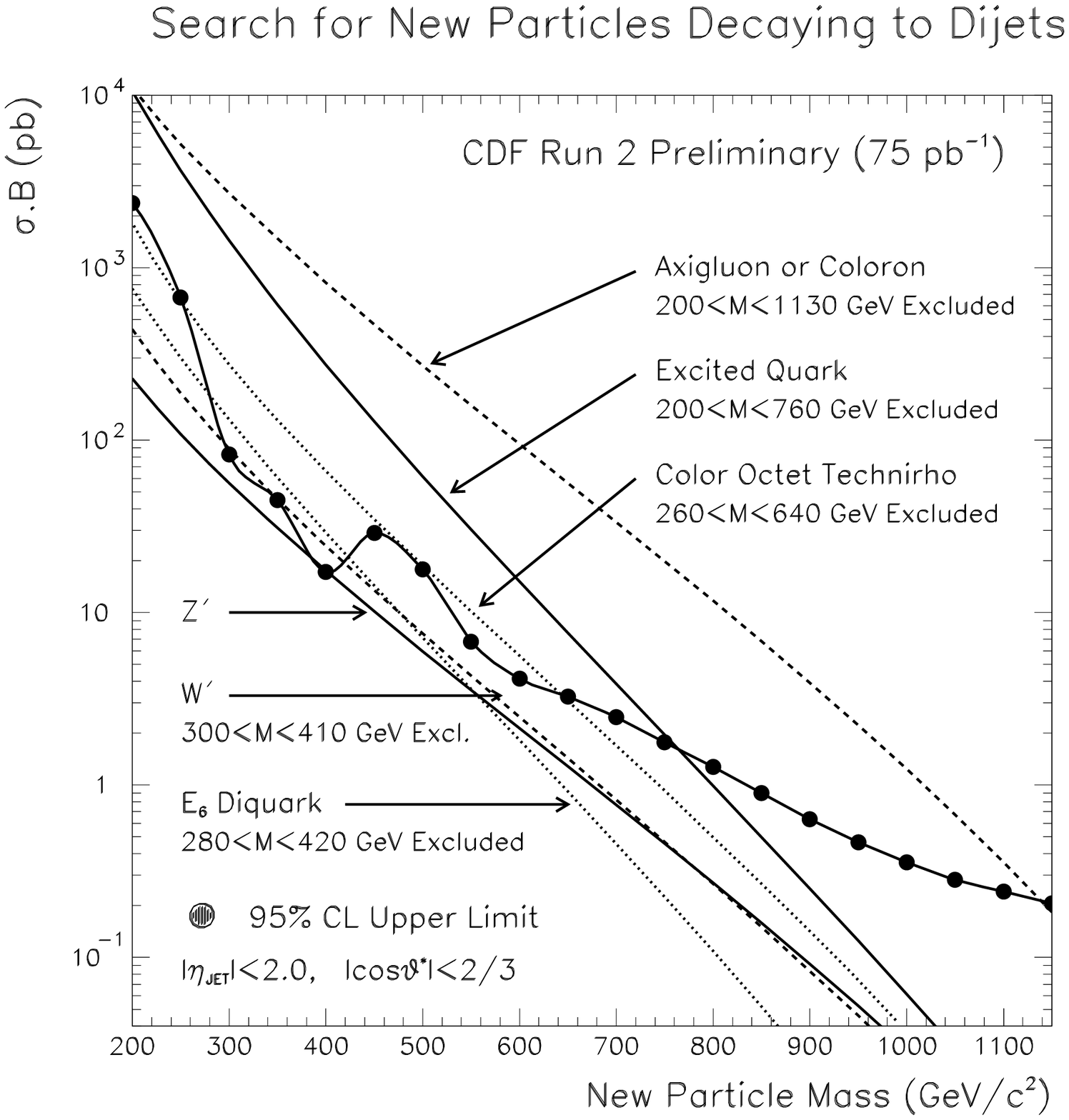,width=6.7cm,height=5.9cm}}
   \end{picture}
   \vspace*{8pt}
   \caption{{\em Left:} Dijet mass spectrum. {\em Right:} 95\% CL upper limits on the cross section
      times branching ratio for narrow dijet resonances, and predictions for axigluons,
      flavor universal colorons, excited quarks, Color Octet Technirhos, E6 diquarks,
       W' and Z'.}
   \label{dijets}
  \end{figure}

  A measurement of the dijet mass spectrum using $75\:\mbox{pb}^{-1}$ of data is
  shown in Fig.~\ref{dijets}, extending from $180\:\mbox{GeV}$ and falling
  steeply, with the event with the highest dijet mass at $1364\:\mbox{GeV}$. The
  ratio of the cross sections measured in Run~I and Run~II compares well with QCD
  calculations, and the Run~II cross section is above the Run~I measurement by a
  factor of 1.1 (low mass) to more than 2 (high mass). Performing a simple
  parameterization of the spectrum, and taking into account the fractional
  difference between the data and the fit as well as the statistical residuals
  between the data and the fit, shows that there is no evidence for a new
  particle compatible with the dijet mass resolution. These data can be used to
  exclude the production of a variety of particles decaying into dijets. The 95\%
  confidence upper limit for cross section times branching ratio together with
  predictions for the production of axigluons, flavor universal colorons, excited
  quarks, Color Octet Technirhos, E6 diquarks, W', and Z' are shown in
  Fig.~\ref{dijets}. All of the exclusion regions are either similar or better
  than the corresponding Run~I limits.

\section{Searches}

  One of the main goals of the Tevatron is to uncover physics beyond the Standard
  Model. A large number of searches is actively pursued, both in the more
  traditional model-based approach, where the analysis is optimized for best
  sensitivity with regard to specific predictions, as well as signature-based
  searches, where the observation of a given event type is compared to Standard
  Model expectations. A few examples are shown in this section.

\begin{figure}[tb]
   \setlength{\unitlength}{1cm}
   \begin{picture}(15.0,4.2)(0.0,0.0)
   \put(-0.1,-0.9){\psfig{file=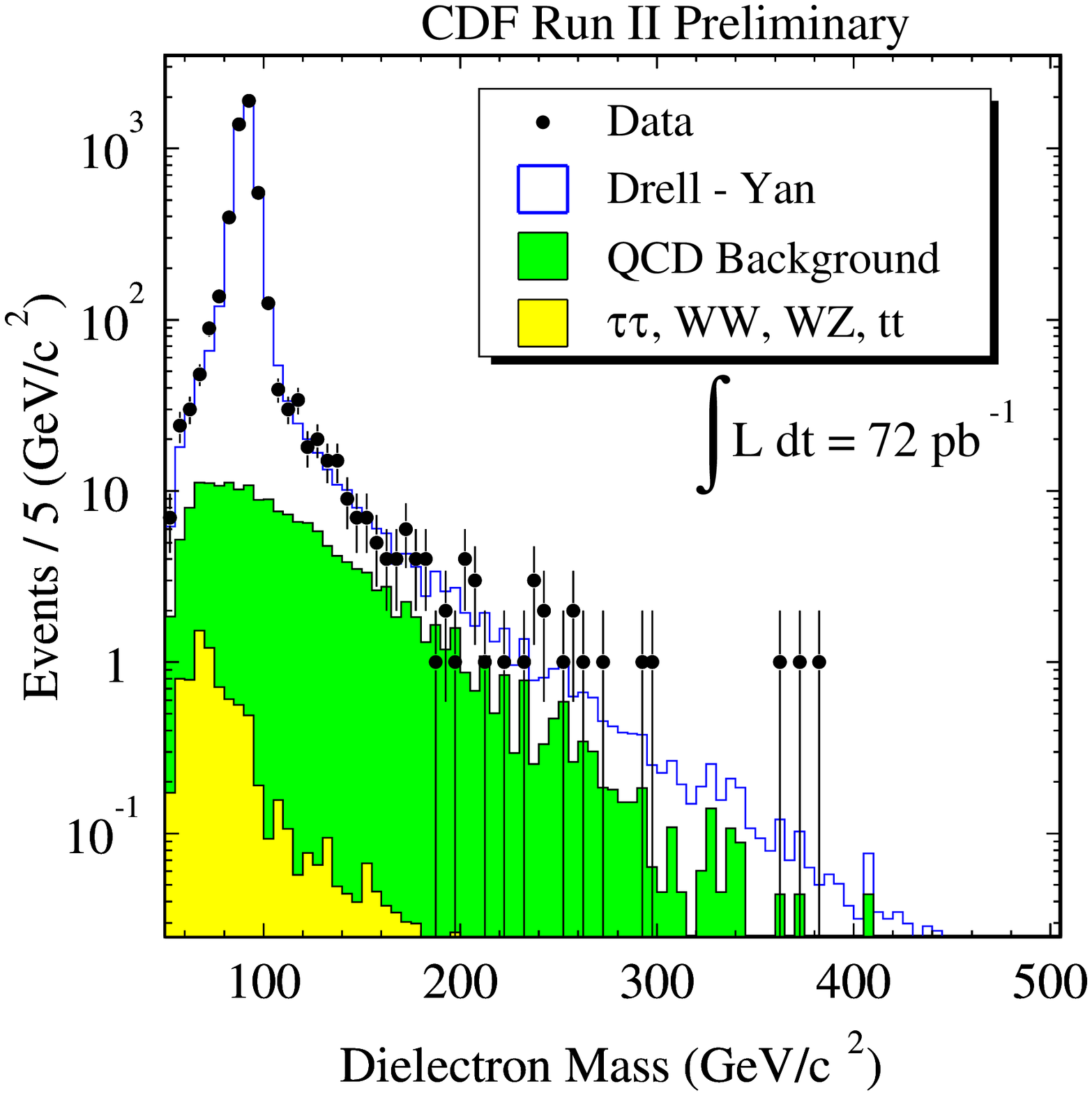,width=6.2cm,height=5.5cm}}
   \put(6.5,-0.65){\psfig{file=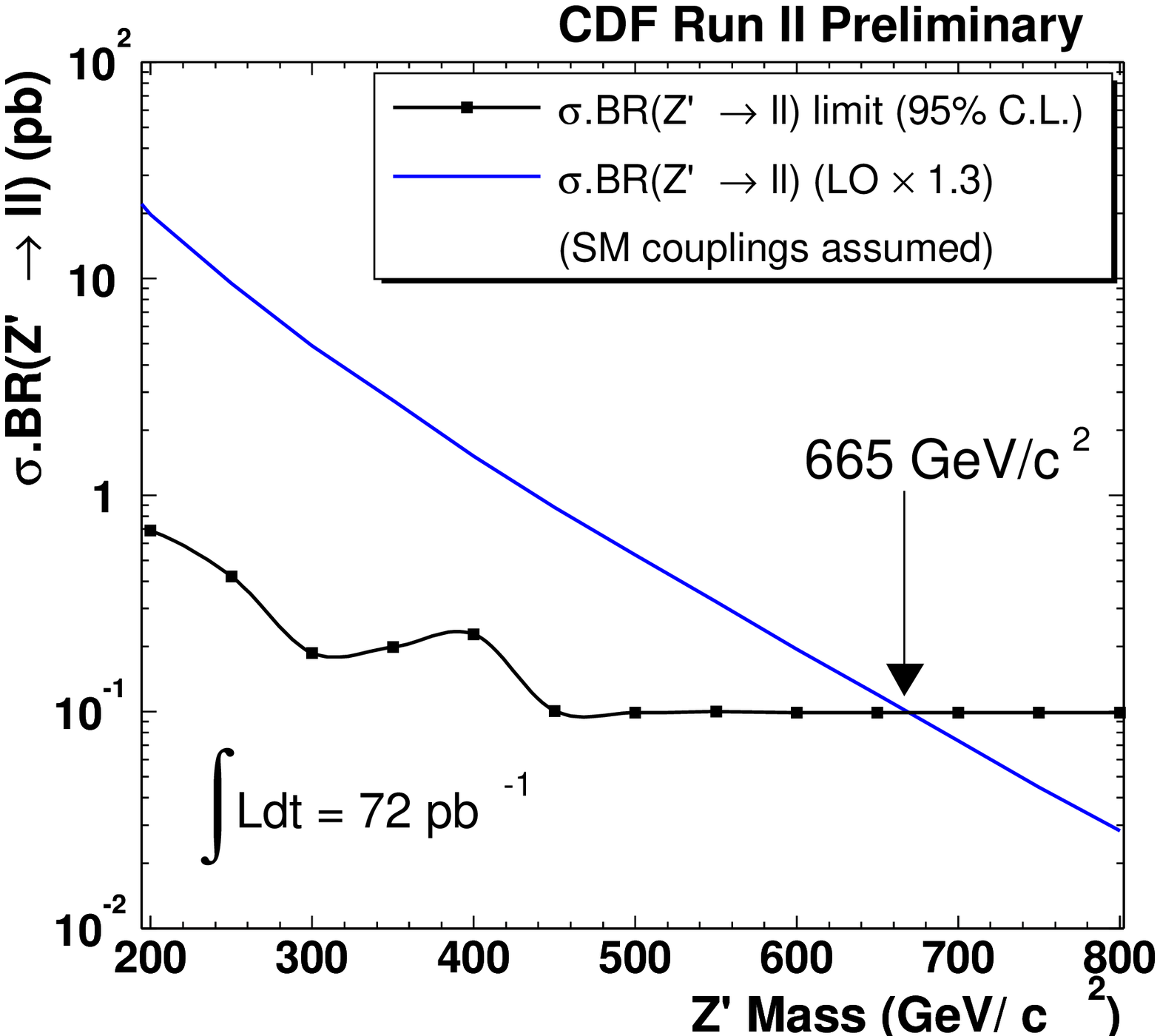,width=6.2cm}}
   \end{picture}
   \vspace*{8pt}
   \caption{{\em Left:} Drell Yan spectrum in the dielectron channel. {\em Right:}
            Limit on $Z'$ production, combining the dielectron and dimuon channels.}
   \label{zprime}
  \end{figure}

  The dilepton mass spectrum was analysed to search for anomalies in the high mass
  region. Additional neutral gauge bosons would show up as resonances in the
  measured dielectron and dimuon mass spectra (see Fig.~\ref{zprime} for the
  dielectron spectrum). The data are well described by Standard Model processes,
  and a limit for the production of a $Z'$ boson is set (Fig.~\ref{zprime}) which
  is comparable to the Run~I analyses. A Randall-Sundrum graviton\cite{rs} would
  show up as a spin 2 resonance in the Drell-Yan spectrum. Limits ranging from
  $205\:\mbox{GeV}$ to $535\:\mbox{GeV}$ are set\cite{rolli}.

\begin{figure}[tb]
   \setlength{\unitlength}{1cm}
   \begin{picture}(15.0,3.4)(0.0,0.0)
   \put(-0.1,-0.8){\psfig{file=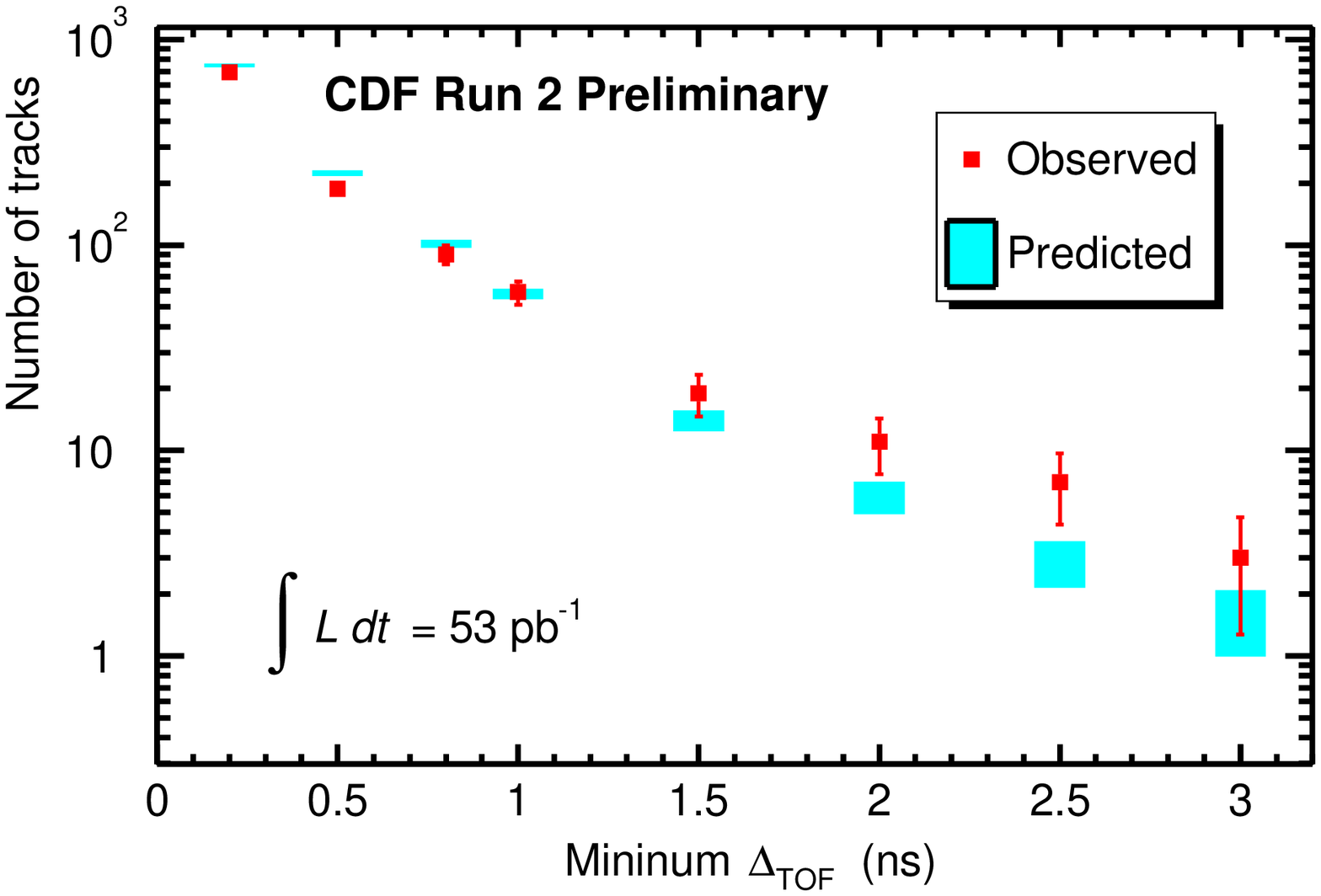,width=6.7cm}}
   \put(6.3,-0.8){\psfig{file=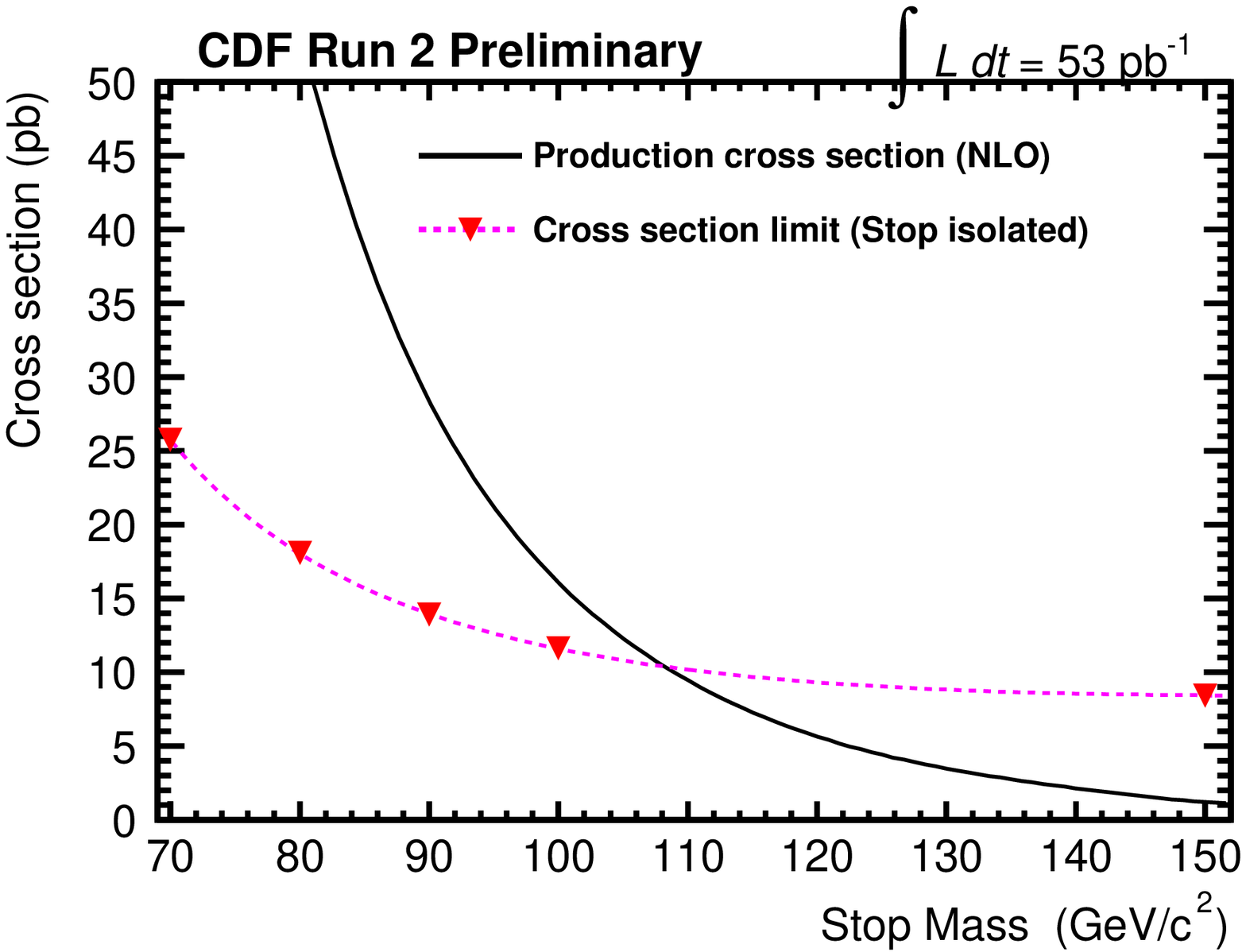,width=6.7cm}}
   \end{picture}
   \vspace*{8pt}
   \caption{Search for CHAMPS. {\em Left:} Observed and predicted number of events as
            function of $\Delta_{TOF}$. {\em Right:} Cross section limit as function of
            stop mass.}
   \label{champs}
  \end{figure}

  An analysis that exploits the new capabilities of the upgraded CDF detector is
  the search for long lived charged massive particles (CHAMPS). These particles
  will have a long time of flight through the detector, and the new TOF system
  provides a sensitivity to higher $\beta\gamma$ than the conventional approach
  of using energy loss ($dE/dx$) measurements provided by the drift chamber. If
  the CHAMP lifetime is long enough to traverse the whole detector, candidate
  events can be collected using the high $p_T$ muon trigger. The discriminant
  variable in the analysis is $\Delta_{TOF}$, which is the difference between
  the measured time of flight for a particle and the time of flight expected for
  a particle travelling at the speed of light. The observed $\Delta_{TOF}$
  spectrum in $53\:\mbox{pb}^{-1}$ of data is shown in Fig.~\ref{champs},
  together with the prediction based on a control sample. For a specific SUSY
  model cross section limits have been set as a function of the stop mass
  (Fig.~\ref{champs}), yielding a mass limit $M_{stop} > 108\:\mbox{GeV}$ at
  95\% confidence level.

\section{Conclusions}

  After a five year shutdown and one year of commissioning, all major components
  of the CDF detector are operating at or near their design specifications for
  the Tevatron Run~II. All relevant benchmark signals are observed and used both
  to characterize the detector performance, and to make several physics
  measurements. In particular, the data collected lately provide insight into
  CDF's heavy flavor capabilities. In spite of the still limited accumulated
  luminosity some measurements already improve the best currently available.

\section*{Acknowledgments}

I wish to thank my colleagues in CDF who provided most of the material.

\end{document}